\documentclass[aps,pra,preprint,superscriptaddress,tightenlines,showpacs]{revtex4}
\usepackage{graphicx} 
\usepackage{amsmath}
 
\newcommand{\beq}{\begin{equation}}
\newcommand{\enq}{\end{equation}}

\begin{document}

\title{Quantum theory of a vortex line in an optical lattice}
\author{J.-P. Martikainen}
\email{J.P.J.Martikainen@phys.uu.nl}
\author{H. T. C. Stoof}
\email{stoof@phys.uu.nl}
\affiliation{Institute for Theoretical Physics, Utrecht University, 
Leuvenlaan 4, 3584 CE Utrecht, The Netherlands}
\date{\today}
\begin{abstract}
We investigate the quantum theory of a vortex line in a stack of 
weakly-coupled two-dimensional Bose-Einstein condensates, that is
created by a one-dimensional optical lattice. We derive the dispersion
relation of the Kelvin modes of the vortex line and also study
the coupling between the Kelvin modes and the quadrupole modes.
We solve the coupled dynamics of the vortex line and the quadrupole modes,
both classically as well as quantum mechanically. The quantum mechanical 
solution reveals the possibility of
generating nonequilibrium squeezed vortex states by strongly driving
the quadrupole modes.
\end{abstract}
\maketitle 

\section{Introduction}
\label{sec:intro}
Quantized vortices play a crucial role in understanding the
stability and decay of superfluid flow. One way to understand this is that
the superfluid
flow of the Bose-Einstein condensate 
in a simply connected region is irrotational. Therefore, the superfluid
can only respond to a rotation by the nucleation of singular topological
defects i.e. vortices. Furthermore, the famous Landau urgument shows that
a superfluid moving past an object
will only experience drag if the superfluid velocity is large enough. 
Again this drag typically manifests itself by the nucleation of vortices.

In the field of ultra cold atomic gases 
a single vortex in a Bose-Einstein condensate was first observed
by Matthews {\it et al.}~\cite{Matthews1999b} and soon after that
also by Madison {\it et al.}~\cite{Madison2000a}. 
Since then the field has developed rapidly. The studies of 
a single or a few 
vortices~\cite{Hodby2001a,Hodby2002a,Rosenbusch2002a,Leanhardt2002a,Leanhardt2003a} 
have more recently also been extended by experiments with
Bose-Einstein condensates with a very large number of 
vortices~\cite{Abo2000a,AboShaeer2002a,Engels2002a,Engels2003a,Coddington2003a}.
Apart from some of the most recent advances,
the review article by Fetter and Svidzinsky~\cite{Fetter2001a}
provides an excellent overview of the extensive 
theoretical and experimental work on vortices in Bose-Einstein condensates.

While vortices are very important for understanding superfluid properties
of liquids and gases,
the research into Bose-Einstein condensates in
optical lattices has provided an important alternative for 
exploring the nature of 
superfluidity. 
Bosons in an optical lattice realize the Bose-Hubbard model~\cite{Jaksch1998a}
of which the ground-state is, depending on parameters, either a superfluid or 
a Mott-insulator~\cite{Jaksch1998a,vanOosten2001a,Roth2003a,vanOosten2003a,Dickerscheid2003a}. 
The quantum phase transition
between these two phases was indeed recently observed in a beautiful experiment 
by Greiner {\it et al.}~\cite{Greiner2002a}. Furthermore, Bose-Einstein condensates
in an optical lattice have been used to study experimentally diffraction of matter 
waves~\cite{Ovchinnikov1999a}, number squeezing~\cite{Orzel2001a}, collapses
and revivals~\cite{Greiner2002b}, 
superfluid dynamics~\cite{Cataliotti2001a,Burger2001a}, as well as 
Bloch oscillations~\cite{Morsch2001a,Cristiani2002a}. 
There is also a number of important theoretical papers on the dynamics and instabilities
of a Bose-Einstein condensate in an optical 
lattices~\cite{Trombettoni2001b,Wu2001a,Konotop2002a,Baizakov2002a,
Kraemer2002a,Machholm2003a,Massignan2003a}.

In this paper, we study the quantum physics of a vortex line in an 
one-dimensional optical lattice by extending on our 
earlier work~\cite{Martikainen2003b}. This problem is 
interesting, not only because it combines the above discussed 
two different approaches to the study
of superfluidity in a Bose-Einstein condensate, but also because the quantum properties
of the vortex are expected to be more pronounced in a system with a smaller
number of particles and in lower dimensions.
Furthermore, a Bose-Einstein condensate in a one-dimensional optical lattice
has an intriguingly similar layered structure as the high-$T_c$ superconductors and 
it also appears to be a promising way to reach the quantum 
Hall regime~\cite{Cornell2002private}, for which we
 need to have about one vortex per particle.
Using a variational ansatz for the condensate wave function in each 
site of the optical potential,
we derive from first principles the quantum theory of the vortex line in an 
one-dimensional optical lattice
and obtain after linearization the dispersion relation of the Kelvin modes of the
vortex line. Physically these modes correspond to a sinusoidal
wiggling of the vortex line. 
It turns out that in an optical lattice, as compared
to the system without a periodic potential, the theoretical treatment 
of the vortex line is considerably simplified and 
the dispersion relation can be obtained analytically.

Kelvin modes or kelvons were 
observed for the first time in a magnetically trapped Bose-Einstein condensate 
by Bretin {\it et al.}~\cite{Bretin2003a}
and a theoretical understanding of the observed phenomena 
was provided by Mizushima {\it et al.}~\cite{Mizushima2003a}. In this experiment,
the quadrupole mode with azimuthal quantum number $m=-2$ was observed to excite 
the Kelvin modes. Motivated by this experiment we also use our theory to study
the coupling between the kelvons and the quadrupole modes in a one-dimensional
optical lattice. We find that the coupling is described by 
a squeezing Hamiltonian~\cite{Scully1997a}. This implies that, when
the vortex line is treated quantum mechanically, the quadrupole mode
will generate nonequilibrium squeezed states of the vortex line. 
In Ref.~\cite{Martikainen2003d}
we have discussed also the spontaneous squeezing of the vortex that occurs
in equilibrium if also
interactions between the kelvons are taken into account.
In this paper, however, we neglect these interactions and discuss both 
the classical as well as the quantum dynamics of the vortex line
when the vortex is driven by a strongly excited quadrupole mode. Furthermore,
we discuss the dynamics of vortex squeezing and how to observe 
the resulting squeezed vortex
states experimentally.

The paper is organized as follows. In Sec.~\ref{sec:model} we present
the theoretical foundations of our work. In Sec.~\ref{sec:varansatz}
we discuss our variational approach and derive the Lagrangian for our
system. We apply the results of this section in
Sec.~\ref{sec:eigenmodes}, where we quantize the vortex line and find also 
the dispersion relation of the kelvons. In addition, 
we derive in Sec.~\ref{sec:eigenmodes} 
the quantum theory of the quadrupole modes and solve their
dispersion relations. In Sec.~\ref{sec:coupling} we study the coupling between
the quadrupole modes and the kelvons, and in Secs.~\ref{sec:classicaldyn} 
and~\ref{sec:quantumdyn} we explore, respectively, the classical and quantum dynamics 
of the vortex line that results from this coupling. 
In Sec~\ref{sec:squeezing} we discuss
the nonequilibrium squeezing of the vortex line. We end with a short discussion
of our results in Sec.~\ref{sec:conclusions}.

\section{Energy functional}
\label{sec:model}
Our starting point is a cigar-shaped Bose-Einstein condensate trapped by
the potential
\beq
\label{magnetic_trap}
V({\bf r})=\frac{m}{2}\left(\omega_r^2r^2+\omega_z^2z^2
\right),
\enq
where $\omega_r$ and $\omega_z$ are the radial and axial 
trapping frequencies, respectively, and $m$ is
the atomic mass. As we assume a 
cigar-shaped trap, we further have that $\omega_z\ll \omega_r$. 
The Bose-Einstein condensate also experiences an one-dimensional optical lattice
\beq
\label{lattice_potential}
V_0({\bf r})=V_0\sin^2\left(\frac{2\pi z}{\lambda}\right),
\enq
where $V_0$ is the lattice depth and $\lambda$ is the wave length of the
laser light. We assume that the lattice is deep enough so that it dominates
over the magnetic trapping potential in the $z$-direction. 
When this is true and the number of lattice sites is very large, 
i.e., $\lambda\ll l_z=\sqrt{\hbar/m\omega_z}$,
we can in first instance
ignore the magnetic trapping potential in the $z$-direction.

The lattice potential splits the condensate into $N_s$
two-dimensional condensates with a pancake shape.
We take the lattice to be sufficiently deep
such that its depth is larger than the chemical potential
of the two-dimensional condensate. 
Although we are thus interested in a deep lattice, we consider here
only the case that there is still full coherence 
across the condensate array. Specifically this means that
the lattice potential should not be so deep as to induce a Mott-insulator
transition. Typically the required lattice depth to reach
the Mott-insulator transition in a three-dimensional lattice with 
a filling factor of one is
of the order of $10 E_r$, where $E_r$ is the recoil energy of an atom
after absorbing a photon from the laser beam. 
In a one-dimensional lattice
the number of atoms in each lattice site is typically much larger 
than in a three-dimensional lattice and the transition into 
the insulating state requires a much deeper lattice~\cite{vanOosten2003a}. 
The Mott-insulator transition in a one-dimensional optical lattice
has very recently been observed~\cite{Stoferle2003a}, but also in that case
the filling fraction is of the order of one.

We use trap units from now on, i.e., the unit of energy is $\hbar\omega_r$,
 the unit of time is $1/\omega_r$,
and the unit of length is $l_r=\sqrt{\hbar/m\omega_r}$.
The Gross-Pitaevskii energy functional, which describes the gas
at low temperatures rotating at frequency $\Omega$, is then
\begin{eqnarray}
\label{fullH}
E\left[\Psi^*,\Psi\right]&=&
\int d{\bf r} \Psi^*({\bf r})
\left[-\frac{\nabla^2}{2} +\frac{1}{2}\left(x^2+y^2\right)+\frac{V_0({\bf r})}{\hbar\omega_r}
+\frac{T^{2B}}{2}|\Psi({\bf r})|^2-\Omega \hat{L}_z\right]\Psi({\bf r}),
\end{eqnarray}
where $T^{2B}$ is the two-body $T$-matrix. In the above units the latter is
related to the three-dimensional 
$s$-wave scattering length $a$ through $T^{2B}=4\pi a/l_r$.
In addition, 
\beq
\hat{L}_z=\frac{1}{i}\left(x\frac{\partial}{\partial y}-y\frac{\partial}{\partial x}\right)
\enq
is the angular momentum component in the $z$-direction.

For a deep lattice potential it is natural to expand 
the Bose-Einstein condensate wave function in terms of Wannier functions that
are well localized in the sites. More precisely, we expand
the Bose-Einstein condensate wave function as
\beq
\label{localized_ansatz}
\Psi\left({\bf r}\right)=\sum_n \Phi\left(z-z_{n}\right)
\Phi_n\left(x,y\right),
\enq
where $n$ labels the lattice sites and $z_n=\lambda n/2 l_r$ 
is the position of the
$n$th site. For now we do not specify the 
wave functions $\Phi_n\left(x,y\right)$ of the
two-dimensional condensates, but for the Wannier function
in the $z$-direction, $\Phi\left(z\right)$, we use the ground-state
wave function of the harmonic approximation to the lattice potential near the
lattice minimum. This harmonic trap has the frequency 
\beq
\omega_L=\frac{2\pi}{\lambda}\sqrt{2V_0/m}
\enq
and the wave function $\Phi(z)$ is thus given by
\beq
\label{z_profile}
	\Phi(z)=\frac{1}{\pi^{1/4}\sqrt{l_L}}\; 
\exp\left(-\frac{z^2}{2 l_L^2}\right),
\enq
where $l_L=\sqrt{\hbar/m\omega_L}$.

Substituting the above ansatz into the energy functional
and ignoring all but the nearest-neighbour interactions, we get the 
energy functional
\begin{eqnarray}
\label{2D_Hamiltonian}
E\left[\Phi^*,\Phi\right]
&=&\int dx dy\left\{\sum_n\Phi_n^*(x,y)\left[
-\frac{\nabla^2}{2}
+\frac{1}{2}\left(x^2+y^2\right)+\frac{U_{2D}}{2}|\Phi_n(x,y)|^2
-\Omega\hat{L}_z\right]\Phi_n(x,y)\right.\nonumber\\
&-&\left.J\sum_{<n,m>} \Phi^*_m(x,y)\Phi_n(x,y)\right\},
\end{eqnarray}
where $\langle n,m \rangle$ indicates nearest neighbours, and
\beq
U_{2D}=T^{2B}\int dz |\Phi(z)|^4=4\sqrt{\frac{\pi}{2}}
\left(\frac{a}{l_L}\right)
\enq
is the effective two-dimensional coupling strength. Moreover, $J$ 
is the strength of the Josephson coupling between neighbouring sites
and we have
\begin{eqnarray}
\label{Jdefinition}
J=-\int dz \Phi^*(z)\left[
-\frac{1}{2}\frac{\partial^2}{\partial z^2}+\frac{V_0(z)}{\hbar\omega_r}
\right]
\Phi(z+\lambda/2\,l_r).
\end{eqnarray}
With these assumptions $J$ is a time-independent 
experimentally tunable parameter. 
Approximating the lattice potential near its maximum by an upside-down 
parabolic potential, we can
calculate the Gaussian integral, with the result
\beq
J=\frac{1}{8\pi^2}\left(\frac{\omega_L}{\omega_r}\right)^2
\left(\frac{\lambda}{l_r}\right)^2\left[\frac{\pi^2}{4}-1\right]
e^{-\left(\lambda/4\,l_L\right)^2}.
\enq

The energy functional in Eq.~(\ref{2D_Hamiltonian}) 
is now almost two-dimensional. The third dimension
is visible only in the last term that describes the coupling between
neighbouring layers. The energy is characterised by two 
parameters $U_{2D}$ and $J$, both of which are experimentally tunable.
The importance of the on-site interaction 
term proportional to $U_{2D}$ can be enhanced by increasing 
the number of particles in the sites or by making the lattice deeper.
Deepening the lattice also decreases the strength of the Josephson coupling 
$J$ and makes the sites
more independent. It should be noticed that while $J$ is tunable, it is
always positive. Physically this means that there is always
an energetic penalty for having a phase difference between sites.

\section{Variational ansatz}
\label{sec:varansatz}
We intend to study the coupled system of a displaced vortex line
and the quadrupole modes. Therefore,
a variational ansatz for the two-dimensional Bose-Einstein
condensate wave function $\Phi_n(x,y,t)$ must satisfy two
obvious physical requirements.
Namely, the ansatz for the two-dimensional wave functions in each site
must predict the vortex precession dynamics and the quadrupole mode
frequencies with sufficient accuracy. The ansatz we used in 
an earlier publication~\cite{Martikainen2003a} predicts the quadrupole
frequencies very well from the non-interacting to the strongly-interacting
limit, but unfortunately fails to predict 
in the strongly-interacting or Thomas-Fermi limit
the correct vortex-precession
dynamics. This has to do with the fact that our previous ansatz
does not correctly account for the size of the vortex core in that limit. 
This problem is not important for the collective modes, but turns out to 
have serious adverse effects in predicting the vortex dynamics.
Therefore, we here use the ansatz
\beq
\Phi_n\left(x,y,t\right)\propto \exp\left(-\frac{B_0}{2}\left(x^2+y^2\right)
-\frac{\epsilon_n(t)}{2}\left(x^2-y^2\right)-\epsilon_{xy,n}(t)xy
+i\tan^{-1}\left(\frac{y-y_n(t)}{x-x_n(t)}\right)\right),
\label{ansatz}
\enq
where $(x_n(t),y_n(t))$ is the position of the vortex in the $n$th site and the
variational parameters $\epsilon_n(t)=\epsilon_n'(t)+i\epsilon_n''(t)$ 
and  $\epsilon_{xy,n}(t)=\epsilon_{xy,n}'(t)+i\epsilon_{xy,n}''(t)$,
that describe the quadrupole and scissors modes, are complex. 
In addition, the wave function in each layer is normalized to $N$.
Since our ansatz does not contain a vortex core, the average kinetic energy would diverge
in this case.
Therefore, we introduce a small distance cut-off $\xi$ as an additional
variational parameter. Together with the condensate size $1/\sqrt{B_0}$ 
this parameter will be determined
by minimizing the equilibrium energy functional. 

\subsection{Equilibrium solution}
In equilibrium the system has a straight vortex line
in the center of the condensate. The condensate energy per particle is then
given by
\beq
\frac{E\left(B_0,\xi\right)}{N}=
\frac{1}{2B_0}+\frac{B_0}{2}+B_0U+\frac{\xi^2}{2B_0}
-\frac{B_0^2\xi^2}{2}+\frac{1}{2}B_0e^{B_0\xi^2}\Gamma\left[0,B_0\xi^2\right]
-\Omega,
\label{equi_energy}
\enq
where 
\beq
U=\frac{N}{\sqrt{2\pi}}\left(\frac{a}{l_r}\right)
\sqrt{\frac{\omega_L}{\omega_r}}.
\enq
Furthermore, $\Gamma\left[a,z\right]$ is the incomplete gamma function.
Eq.~(\ref{equi_energy}) 
is easy to minimize numerically for both $\xi$ and $B_0$, but 
expressions later on simplify when we
note that the short disctance cut-off is given, with realistic parameter values, 
to a very good accuracy by $\xi=\sqrt{B_0}$.
After this simplification we still have to minimize the resulting
transcendental equation for the remaining parameter $B_0$. This is a simple
numerical task that must be performed only once in the beginning of the
calculations.

The equilibrium solution reveals two physically obvious features. First,
the Bose-Einstein condensate with the vortex is slightly bigger than 
the Bose-Einstein condensate
without the vortex. This is due to the centrifugal potential
arising from the vortex velocity pattern.
Second, the size of the vortex core is smaller
for larger condensates. Also, it is good to keep in mind that
with positive scattering lengths $0<B_0<1$ and that in the
Thomas-Fermi limit of large interactions we have $B_0\ll 1$.
The ansatz in Eq.~(\ref{ansatz}) provides an accurate description
of the system when the interaction strength $U$ is larger than about $10$.
The ansatz fails for very weakly-interacting 
Bose-Einstein condensates, as then
the energy in Eq.~(\ref{equi_energy}) ceases to have a minimum.
In this regime our previous ansatz can however
be used~\cite{Martikainen2003a}.

\subsection{Second-order expansion of the Lagrangian}
As we are interested in the kelvon and quadrupole excitations we must
expand the system's Lagrangian to second order in $\epsilon$,
$\epsilon_{xy}$, $x_n$, and $y_n$. To second order there is no
coupling between the quadrupole modes and the vortex displacements.
Consequently, we can treat them independently. Also,
as we are expanding around the equilibrium solution,
we can safely ignore the zeroth and first-order contributions.
The Lagrangian $L_\alpha=T_\alpha-E_\alpha$, with $\alpha=Q\,{\rm or}\, K$, for the
quadrupole modes and the kelvons contain also the
contributions $T_\alpha$ that result from the general time-derivative term 
in the action for the Bose-Einstein condensate given by
\beq
T\left[\Phi^*,\Phi\right]=
\frac{i}{2}\sum_n\int dxdy \left[\Phi_n^*\dot{\Phi}_n-
\Phi_n\dot{\Phi}_n^*\right],
\enq
as well as the energy contribution $E_\alpha$
that follow from $E\left[\Psi^*,\Psi\right]$. For the
quadrupole modes we obtain
\beq
\frac{T_Q}{N}=-\frac{1}{2B_0^2}\sum_n \left(\epsilon_n'\dot{\epsilon_n}''
+\epsilon_{xy,n}'\dot{\epsilon}_{xy,n}''\right)
\enq
and
\begin{eqnarray}
\frac{E_Q}{N}&=&\sum_n \left(\frac{1}{2B_0^3}-\frac{U}{2B_0}+\frac{1}{8B_0}
-\frac{\Gamma\left[0,B_0^2\right]}{4B_0}\right)
\left(\epsilon_{n}'^2+\epsilon_{xy,n}'^2\right)
+\frac{1}{2B_0}\left(\epsilon_n''^2+\epsilon_{xy,n}''^2\right)
\nonumber\\
&+&\left(\frac{1}{2B_0}-\frac{\Omega}{B_0^2}\right)
\left(\epsilon_n'\epsilon_{xy,n}''
-\epsilon_n''\epsilon_{xy,n}'\right)
+\frac{J}{8B_0^2}\sum_{\langle n,m\rangle}
\left[\left(\epsilon_{n}'-\epsilon_{m}'\right)^2
+\left(\epsilon_{n}''-\epsilon_{m}''\right)^2
\right.\nonumber\\
&+&\left.\left(\epsilon_{xy,n}'-\epsilon_{xy,m}'\right)^2
+\left(\epsilon_{xy,n}''-\epsilon_{xy,m}''\right)^2
\right].
\end{eqnarray}
For the kelvons we get,
in the lowest non-vanishing order in $B_0$, that
\beq
\frac{T_K}{N}=B_0\sum_n \left(y_n\dot{x}_n-x_n\dot{y}_n\right)
\label{TK}
\enq
and
\begin{eqnarray}
\frac{E_K}{N}&=&\frac{B_0^2}{2}\sum_n \left(1-\Gamma\left[0,B_0^2\right]
+\frac{2\Omega}{B_0^2}\right)
\left(x_n^2+y_n^2\right)\nonumber \\
&+&\frac{JB_0\Gamma\left[0,B_0^2\right]}{4}
\sum_{\langle n,m\rangle}\left[\left(x_n-x_m\right)^2+
\left(y_n-y_m\right)^2\right].
\label{EK}
\end{eqnarray}
In the absence
of Josephson coupling the equations of motion for the vortex position
describe the precession of a displaced vortex around the center of
the condensate. With realistic interaction
strengths the vortex-precession frequency predicted by
Eqs.~(\ref{TK}) and~(\ref{EK})
is close to the result
using a Thomas-Fermi wave function for the disk-shaped 
condensate~\cite{Fetter2001a}.

\section{Eigenmodes of the system}
\label{sec:eigenmodes}
With the second-order expansion of the Lagrangian
available, we now proceed to study the eigenmodes of the system.
Our variational ansatz in Eq.~(\ref{ansatz}) captures three
eigenmodes, namely
kelvons and two quadrupole modes with quantum numbers
$m=\pm 2$. In the next two 
sections we solve the kelvon and the quadrupole mode dispersion relations.
We deal with the nearest-neighbour 
Josephson coupling by transforming to 
momentum space in the same way as in our previous 
work~\cite{Martikainen2003a}.
In momentum space the terms due to the Josephson coupling 
are diagonal and are not more complicated than the other 
contributions in the Lagrangian.

\subsection{Kelvon dispersion relation}
\label{sec:kelvondispersion}
From now on we choose to interpret vortex displacements as
quantum mechanical operators. This involves the important but natural
approximation that the field operator $\hat{\psi}_n\left(x,y,t\right)$, 
which annihilates a boson in the $n$th layer in position $(x,y)$ at time $t$, 
can be replaced with $\Phi_n\left(x,y,\hat{x}_n(t),\hat{y}_n(t)\right)$. 
Physically this implies that the system can be well described by
a classical condensate wave function $\Phi_n\left(x,y\right)$ and that
all the important quantum fluctations are in the vortex position.
Using the results in the previous
section and transforming to momentum space using the convention
$f_k=1/\sqrt{N_s}\,\sum_n e^{-ikz_n} f_n$,
we find that  
in momentum space the Lagrangian for the vortex line becomes
\begin{eqnarray}
\frac{\hat{L}_K}{N}&=&B_0\sum_k \left(\hat{y}_k\dot{\hat{x}}_{-k}-
\hat{x}_k\dot{\hat{y}}_{-k}\right)
\nonumber\\ 
&-&B_0\sum_k \left[\frac{B_0}{2}-\frac{B_0\Gamma\left[0,B_0^2\right]}{2}
+\Omega
+J(k)\Gamma\left[0,B_0^2\right]\right]\left(\hat{x}_k\hat{x}_{-k}+\hat{y}_k\hat{y}_{-k}\right)
\end{eqnarray}
where $J(k)=J\left[1-\cos(kd)\right]$ and $d=\lambda/2$. 
We proceed by defining
the vortex positions in terms of bosonic creation and annihilation
operators as $\hat{x}_k=(\hat{v}_{-k}^\dagger+\hat{v}_k)/2\sqrt{NB_0}$ and 
$\hat{y}_k=i(\hat{v}_{-k}^\dagger-\hat{v}_k)/2\sqrt{NB_0}$~\cite{Fetter1967a}.
In this manner the Lagrangian takes the desired form
\beq
\hat{L}_K=\sum_k \left(i\hat{v}_{k}^\dagger\dot{\hat{v}}_k-
\omega_K\left(k\right)\hat{v}_{k}^\dagger\hat{v}_k\right),
\enq
where the kelvon dispersion is given by
\beq
\omega_K(k)=B_0/2+\Omega+\Gamma\left[0,B_0^2\right]\left(2J(k)-B_0/2\right).
\enq
The solution at zero momentum corresponds to the precession frequency
of a straight but slightly displaced vortex line.

It is important to realize that due to the Euler dynamics of the vortex, the
coordinates $\hat{x}_n$ and $\hat{y}_n$ are canonically conjugate variables
and obey the equal-time Heisenberg commutation relation
\beq
\left[\hat{x}_n,\hat{y}_n\right]=\frac{i}{2NB_0}.
\enq
Our theory is valid if the vortex line precesses well within the condensate. 
For the straight vortex line this condition implies that 
the average displacement $\sqrt{\langle \hat{x}_n^2\rangle+
\langle \hat{y}_n^2\rangle}$ should be much smaller than $1/\sqrt{B_0}$.
This leads to a limitation for the expectation value of the 
number operator for the kelvons, namely,
\beq
\frac{1}{N_s}\sum_k\langle \hat{v}_{k}^\dagger\hat{v}_{k}\rangle
\ll N.
\enq
Consequently, the total number of condensate atoms
sets an upper bound for the kelvon numbers.

Our result for the kelvon dispersion relation has some interesting 
properties. First, the initially negative precession frequency can change sign
if the momentum $k$ of the kelvon is large enough. 
Moreover, the dispersion is quadratic for small momenta. 
Both features are due to
the attractive interactions between neighbouring pancake 
vortices, which is harmonic
for small separations. 
This contrasts with the behaviour of the vortex line in a three-dimensional
bulk superfluid, where the kelvon dispersion 
relation behaves like $k^2 \ln\left(1/k\xi\right)$ and
has a logarithmic dependence for small $k$~\cite{Fetter1967a,Fetter2004a}.
Third, the vortex position is ``smeared'' by the quantum fluctuations. 
In particular we have for the kelvon vacuum state that 
$\langle \hat{x}_n^2\rangle=\langle \hat{y}_n^2\rangle=1/4NB_0$. 
Therefore, the quantum properties of the vortex 
become more important in a lattice. This is due to
the reduced particle number in every site, 
as opposed to the total number of particles,
and the spreading out of the condensate wave function
as the lattice depth is increased.

\subsection{Quadrupole mode dispersion relations}
\label{sec:quadrupoledispersion}
The quadrupole modes are technically somewhat more complicated than
the kelvons, but their dispersion relations can be derived
in a similar way. 
The quadrupole modes are characterized by the 
$4$-component vector $\left(\hat{\epsilon}_k' \;, \hat{\epsilon}_k'' \;, 
\hat{\epsilon}_{xy,k}' \;, \hat{\epsilon}_{xy,k}''\right)$. By writing down
equations of motion for all $4$ variational parameters,
we get a set of coupled first-order differential equations.
The problem of finding the eigenmodes is then reduced to the problem 
of finding the eigenvectors of a $4\times 4$ matrix. In our
case this can be done analytically. We find four eigenvectors
\beq
{\bf v}_1=\left(\begin{array}{c}
A(k)\\-i\\i A(k)\\1
\end{array}
\right),
\enq
\beq
{\bf v}_3=\left(\begin{array}{c}
-A(k)\\i\\iA(k)\\1
\end{array}
\right),
\enq
${\bf v}_2={\bf v}_1^*$, and ${\bf v}_4={\bf v}_3^*$.
Of these solutions
${\bf v}_1$ and ${\bf v}_2$ represent the quadrupole mode with azimuthal 
quantum number $m=2$, and
${\bf v}_3$ and ${\bf v}_4$ represent the quadrupole mode with azimuthal 
quantum number $m=-2$.
The function $A(k)$ is given by
\beq
A\left(k\right)=\sqrt{\frac{4B_0^2+8B_0J(k)}{4+8B_0J(k)+
B_0^2\left(1-4U-2\Gamma\left[0,B_0^2\right]\right)}}.
\enq

We can now expand any vector 
$\left(\hat{\epsilon}_k' \;, \hat{\epsilon}_k'' \;,
\hat{\epsilon}_{xy,k}' \;, \hat{\epsilon}_{xy,k}''\right)$ into this basis
of eigenvectors.
Since $\left(\hat{\epsilon}_k' \;, \hat{\epsilon}_k'' \;, 
\hat{\epsilon}_{xy,k}' \;, \hat{\epsilon}_{xy,k}''\right)$
is a vector of hermitian operators, it is sufficient
to deal only with the complex expansion amplitudes $\hat{a}_1$ and 
$\hat{a}_3$ of the eigenvectors ${\bf v}_1$ and ${\bf v}_3$
respectively. When we define the quadrupole creation and
annihilation operators with
\beq
\hat{a}_1=\sqrt{\frac{2}{NA(k)}}\frac{B_0}{4}\left(\hat{q}_{2,-k}^\dagger+
\hat{q}_{2,k}
+i\,\hat{q}_{2,-k}^\dagger-
i\,\hat{q}_{2,k}\right)
\enq
and
\beq
\hat{a}_3=\sqrt{\frac{2}{NA(k)}}\frac{B_0}{4}\left(\hat{q}_{-2,-k}^\dagger+
\hat{q}_{-2,k}
+i\,\hat{q}_{-2,-k}^\dagger-
i\,\hat{q}_{-2,k}\right),
\enq
we ultimately find the desired Lagrangian for the quadrupole modes 
\beq
\hat{L}_Q=\sum_{m=\pm 2}\sum_k \left(i\hat{q}_{m,k}^\dagger\dot{\hat{q}}_{m,k}
-\omega_m\left(k\right)\hat{q}_{m,k}^\dagger\hat{q}_{m,k}\right),
\enq
where the quadrupole-mode frequencies are then given by
\begin{eqnarray}
\omega_{\pm 2}\left(k\right)&=&\frac{B_0}{A\left(k\right)}+
A\left(k\right)\left(\frac{1}{B_0}+\frac{B_0}{4}
-UB_0-\frac{B_0\Gamma\left[0,B_0^2\right]}{2}\right)
\nonumber\\
&+&
2J(k)\left(A\left(k\right)+\frac{1}{A\left(k\right)}\right)
\pm \left(B_0-2\Omega\right).
\label{quadrupolefrequencies}
\end{eqnarray}
This expression for the quadrupole frequencies 
generalizes our earlier result in Ref.~\cite{Martikainen2003b}
to the case of a rotating trap with $\Omega\neq 0$.

If we are to consider the quadrupole modes as a small
disturbance, the distortion of the condensate due to the 
quadrupole modes should not be too
strong. Quantitatively, we require that the energy of the quadrupole modes 
is much smaller than the ground-state energy of the Bose-Einstein 
condensate. In estimating the condensate
energy we can in first instance ignore the small contribution from
the vortex. In this way we obtain an upper limit for the 
expectation values of the quadrupole number operators as 
\beq
\frac{1}{N_s}\sum_k\langle \hat{q}_{\pm 2,k}^\dagger\hat{q}_{\pm 2,k}\rangle\ll
\sqrt{1+2U}.
\enq

\begin{figure}
\includegraphics[width=\columnwidth]{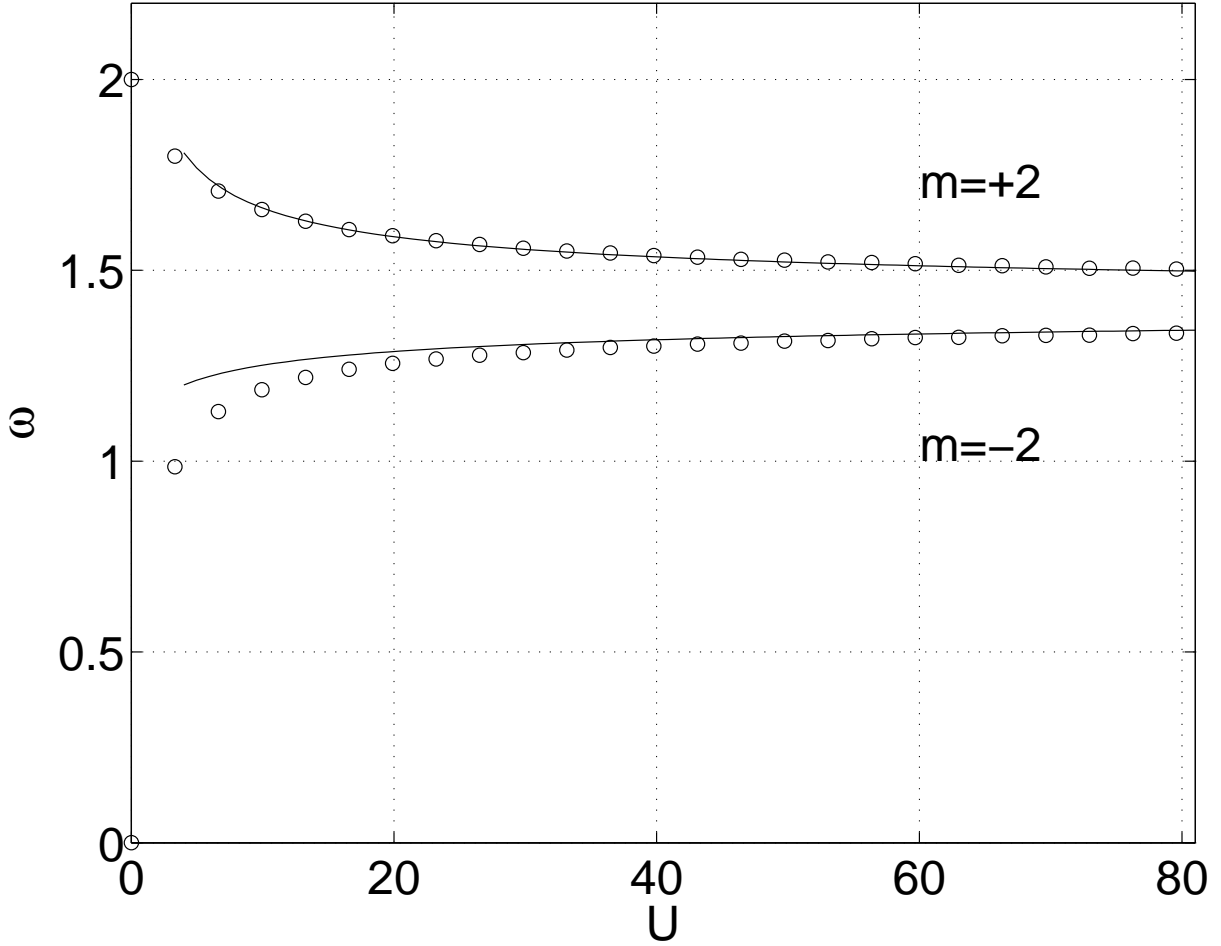}
\caption[]{Frequencies of the $m=\pm 2$ quadrupole modes 
in units of $\omega_r$ for 
$k=0$ and $\Omega=0$ and as a function of interaction strength.
The solid line is calculated using Eq.~(\ref{quadrupolefrequencies}) 
and the open circles are calculated by solving the
appropriate Bogoliubov- de Gennes equations numerically.
\label{fig:quadrupolecomparison}}
\end{figure}

In Fig.~\ref{fig:quadrupolecomparison} we compare the quadrupole
dispersion relations in Eq.~(\ref{quadrupolefrequencies})
for momentum $k=0$ and $\Omega=0$, with the numerical solution of the
Bogoliubov-de Gennes equations for the two-dimensional Bose-Einstein condensate.
As we explained earlier, our ansatz fails
for very weakly-interacting Bose-Einstein condensates. Nevertheless,
Fig.~\ref{fig:quadrupolecomparison} demonstrates
that Eq.~(\ref{quadrupolefrequencies}) 
predicts the quadrupole-mode frequencies well within a wide range
of experimentally relevant interaction strengths.

\section{The coupling between kelvons and quadrupole modes}
\label{sec:coupling}
In the experiment by Bretin {\it et al.}~\cite{Bretin2003a}
the kelvons were excited by the quadrupole mode with a quantum 
number $m=-2$. Therefore,
we now proceed to apply our theory to investigate also the coupling between the
kelvons and the quadrupole modes.
To obtain coupling between the kelvons and the quadrupole modes,
we must expand the Lagrangian up to third order.
At third order we find important contributions from the kinetic
energy, the angular momentum term, as well as from the time-derivative term. 
At third order there is also a coupling between the kelvons and
the quadrupole modes due to the 
Josephson coupling between the sites, but since we assume a small strength
for this coupling we can safely ignore this.
The dominant contributions are respectively given by
\beq
\frac{K^{(3)}}{N}=\frac{B_0\Gamma\left[0,B_0^2\right]}{2}\sum_n\left[
\epsilon_n'\left(y_n^2-x_n^2\right)
-2\epsilon_{xy,n}'x_ny_n
\right]
\enq
\beq
\frac{-\Omega \langle\hat{L_z}\rangle^{(3)}}{N}=
\frac{\Omega}{2}\sum_n\left[
\epsilon_n'\left(x_n^2-y_n^2\right)+2\epsilon_{xy,n}'x_ny_n
\right]
\enq
and
\beq
\frac{T^{(3)}}{N}=\sum_n
\left[-\frac{\epsilon_n'}{2}\left(y_n\dot{x}_n+x_n\,\dot{y}_n\right)+
\frac{\epsilon_{xy,n}'}{2}\left(x_n\dot{x}_n-y_n\dot{y}_n\right)\right].
\enq
To obtain the above relatively simple expression 
for $T^{(3)}$, we expanded
up to first order in $B_0$. Contributions of second-order in
$B_0$ depend on the details of the vortex core profile and are
neglected in the following.

By using our earlier definitions for the kelvon and
quadrupole creation- and annihilation operators 
we can express the third-order contributions to
the energy as
\begin{eqnarray}
\hat{E}^{(3)}&=&
\left(B_0\Gamma\left[0,B_0^2\right]-\Omega\right)
\sum_{k,k'}
\sqrt{\frac{A(k)}{8N_s}}\left[
\hat{q}_{-2,k}\hat{v}_{k+k'}^\dagger\hat{v}_{-k'}^\dagger
+\hat{q}_{-2,-k}^\dagger\hat{v}_{-k-k'}^\dagger\hat{v}_{k'}
\right.\nonumber\\
&-&\left.\hat{q}_{2,k}\hat{v}_{-k-k'}\hat{v}_{k'}
-\hat{q}_{2,-k}^\dagger\hat{v}_{k+k'}^\dagger\hat{v}_{-k'}^\dagger\right]
\end{eqnarray}
and the third order contribution to the time-derivative term as
\begin{eqnarray}
\hat{T}^{(3)}&=&i \sum_{k,k'}
\sqrt{\frac{A(k)}{8N_s}}\left[
\hat{q}_{-2,k}\hat{v}_{k+k'}^\dagger\dot{\hat{v}}_{-k'}^\dagger-
\hat{q}_{-2,-k}^\dagger\hat{v}_{-k-k'}\dot{\hat{v}}_{k'}
\right.\nonumber\\
&+&\left.\hat{q}_{2,k}\hat{v}_{k'}\dot{\hat{v}}_{-k-k'}-
\hat{q}_{2,-k}^\dagger\hat{v}_{k+k'}^\dagger\dot{\hat{v}}_{-k'}^\dagger
\right].
\label{T3_term}
\end{eqnarray}

Using these expressions we can write down and solve the classical
equation motions, but due to the time-derivatives
in Eq.~(\ref{T3_term}) the solution of the operator equations of motion
is difficult. However, it turns out that within a wide range
of experimentally realistic parameter values we can
use the rotating-wave approximation to great effect and thus avoid
the complication introduced by the time derivatives
in Eq.~(\ref{T3_term}). 
We, therefore, introduce the transformation 
$\hat{v}_k\rightarrow \hat{v}_ke^{-i\omega_K(k)t}$
and $\hat{q}_{\pm 2,k}\rightarrow \hat{q}_{\pm 2,k}e^{-i\omega_{\pm 2}(k) t}$
and ignore the remaining small time-dependent term 
in the transformed $\hat{T}^{(3)}$. In this way we obtain the interaction
picture Hamiltonians $\hat{H}_{m}$ for the coupling between the kelvons and the
quadrupole mode with the quantum number $m$ as
\begin{eqnarray}
\hat{H}_{-2}&=&\sum_{k,k'}
\sqrt{\frac{A(k')}{8N_s}}
\left(B_0\Gamma\left[0,B_0^2\right]-\Omega+\omega_K\left(k\right)\right)
\left(\hat{q}_{-2,k'}\hat{v}_{k+k'}^\dagger\hat{v}_{-k}^\dagger
e^{-i\delta_{-2}(k,k')t}\right.\nonumber\\
&+&\left.\hat{q}_{-2,-k'}^\dagger\hat{v}_{-k-k'}\hat{v}_{k}
e^{i\delta_{-2}(k,k')t}
\right)
\end{eqnarray}
and
\begin{eqnarray}
\hat{H}_{2}&=&\sum_{k,k'}\sqrt{\frac{A(k')}{8N_s}}
\left(-B_0\Gamma\left[0,B_0^2\right]+\Omega-\omega_K\left(k\right)\right)
\left(\hat{q}_{2,k'}\hat{v}_{-k-k'}\hat{v}_{k}
e^{-i\delta_{2}(k,k')t}\right.\nonumber\\
&+&\left.
\hat{q}_{2,-k'}^\dagger\hat{v}_{k+k'}^\dagger\hat{v}_{-k}^\dagger\right)
e^{-i\delta_{2}(k,k')t},
\end{eqnarray}
where  
$\delta_{-2}(k,k')=\omega_{-2}(k')-\omega_K(k+k')-\omega_K(k)$
and $\delta_{2}(k,k')=-\omega_{-2}(k')-\omega_K(k+k')-\omega_K(k)$.
The usual experimental preparation will typically imply that only the
quadrupole mode with momentum $k'=0$ is excited.
Therefore, we assume from now on
that only the $k'=0$ quadrupole mode is present and drop
the momentum index from the operator $\hat{q}_{2,-k'}$.

As we show later in Secs.~\ref{sec:classicaldyn}
and~\ref{sec:quantumdyn}, the time evolution of the kelvons
is strongly dominated by the almost resonant modes inside 
a narrow region in momentum space.
Within this region the rotating-wave approximation is clearly appropriate. 
Furthermore, we have solved the classical equations of motion 
numerically without using any approximations.
In this way we verified that the solution given by 
the rotating-wave approximation is indeed very accurate.

\begin{figure}
\includegraphics[width=\columnwidth]{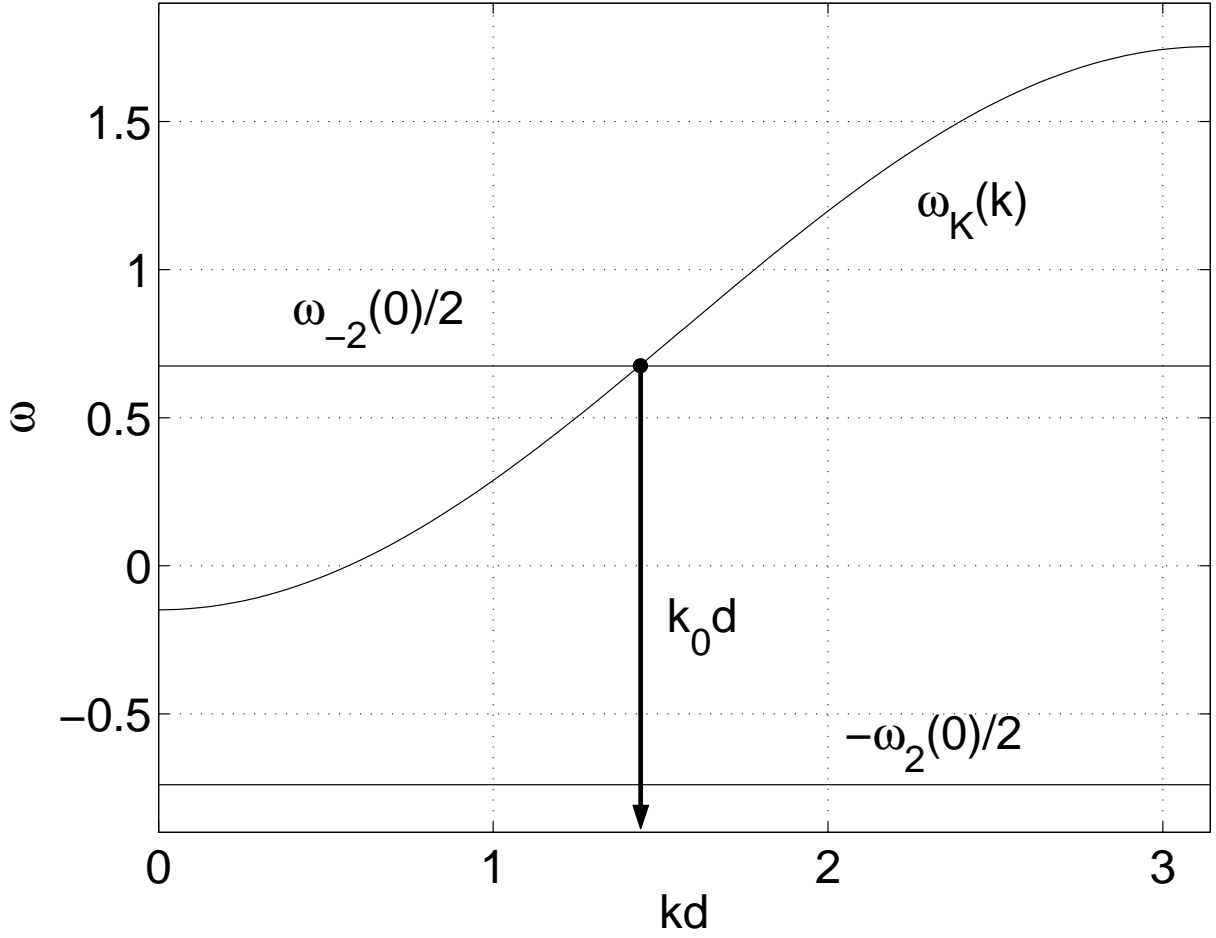}
\caption[]{Demonstration of the resonance condition for the
coupling between the kelvons and the quadrupole mode
with $m=-2$ when $J=0.1$, $U=100$, and $\Omega=0$.
The coupling becomes important when the kelvon
frequency $\omega_K(k)=\omega_{-2}(0)/2$, i.e., when the lines
cross. Since the lowest line does not cross the kelvon dispersion,
the coupling between kelvons and the quadrupole mode
with $m=2$ is very weak. The unit of frequency is $\omega_r$.
\label{fig:resonancedemo}}
\end{figure}

The coupling between the kelvons and quadrupole modes can only
be important if the detunings $\delta_{\pm 2}(k,0)$
are small. Moreover, the coupling is resonant, when the detuning vanishes.
Since in the absence of rotation $\omega_K(0)$ is small and negative and 
$\omega_K(k)$ increases with increasing $k$, whereas $\omega_{\pm 2}(0)$ is positive,
such a resonance is only possible between $m=-2$ quadrupole mode
and the kelvons. We demonstrate this graphically in 
Fig.~\ref{fig:resonancedemo}. This resonance condition
singles out the wavelength $2\pi/k_0$ for the ensuing wiggles in the 
vortex line. The resonance is possible only if the kelvon frequency
increases quickly enough with increasing momentum $k$.
If, for example, the strength of the
Josephson coupling $J$ is too small, the resonance is not possible
at any value of momentum. In order to have a resonant coupling
between the kelvons and the quadrupole mode with $m=-2$ the approximate 
condition
\beq
J\Gamma\left[0,B_0^2\right]>\frac{1}{2\sqrt{2}}
\enq
must be satisfied.
In the absence of rotation
the detuning for the coupling between kelvons
and $m=2$ quadrupole modes is always large and therefore
the rotating-wave approximation eliminates it.

For future convenience we define the
coupling strength between the kelvons and the quadrupole mode 
with $m=-2$ as
\beq
E_c(k,k')=\sqrt{\frac{A(k')}{8N_s}}\left(
B_0\Gamma\left[0,B_0^2\right]-\Omega+\omega_K(k)
\right).
\enq
In terms of this quantity the interaction
picture Hamiltonian becomes
\begin{eqnarray}
\hat{H}_{-2}&=&\sum_{k,k'}
E_c(k,k')
\left(\hat{q}_{-2,k'}\hat{v}_{k+k'}^\dagger\hat{v}_{-k}^\dagger
e^{-i\delta_{-2}(k,k')t}
+\hat{q}_{-2,-k'}^\dagger\hat{v}_{-k-k'}\hat{v}_{k}
e^{i\delta_{-2}(k.k')t}
\right).
\end{eqnarray}
This interaction picture Hamiltonian is similar to the Hamiltonian
encountered in studies of parametric processes in quantum 
optics~\cite{Scully1997a}. Theories describing
the coupling between the atomic and molecular condensate
have also a similar 
structure~\cite{Javanainen1999b,Timmermans1999a,Holland2001a,Duine2003a,
Duine2003b}.
Namely, two atoms (kelvons) are combined to form a single
molecule (quadrupole mode).

While the coupling between the kelvons and $m=-2$ quadrupole modes
is more prevalent, the situation is slightly more involved in
the rotating case. In a rotating trap the frequency of
the quadrupole modes is shifted by $\mp 2\Omega$. For positive
rotation frequencies this shift implies that the 
resonance with quadrupole modes with $m=-2$ and the kelvons
is shifted to a higher
value of momentum or possibly disappears altogether.
On the other hand, the frequency of the quadrupole mode
with $m=2$
is lowered. This signifies that 
with sufficiently large rotation frequencies
the quadrupole mode with $m=2$
can become resonant with the kelvons. In the Thomas-Fermi limit
this implies rotation frequencies larger than about $1/\sqrt{2}$.
At such high rotation frequencies 
the dynamical instability toward the entry of 
additional vortices is likely to play a role~\cite{Madison2001a,Hodby2002a}. 
This unaccounted physical process 
makes the possibility of a strong coupling between the kelvons and
$m=2$ quadrupole modes a technical peculiarity
of our variational ansatz and
renders it, most likely, experimentally irrelevant.

In this paper we focus on the Bose-Einstein
condensate in a one-dimensional
lattice, but it is clear
that the results in this section can also be applied to
study an elongated three-dimensional Bose-Einstein condensate. 
The theory for the coupling between kelvons and
quadrupole modes in such a system is formally similar
to the one we presented here. When 
only quadrupole modes with $k'=0$ are relevant, there are only
two modifications. First, the kelvon dispersion relation should be replaced
with the correct dispersion relation for a vortex line 
in an infinite cylinder with a radial density profile.
Second, the quadrupole-mode frequencies should be replaced 
with the correct quadrupole frequencies for the elongated
Bose-Einstein condensate. For extremely elongated condensates
these frequencies have the same functional form as
Eq.~(\ref{quadrupolefrequencies}), but rather than using
the interaction strength $U$ scaled into two dimensions, we must
use the correct three dimensional interaction strength.

\section{Classical dynamics of the vortex line}
\label{sec:classicaldyn}
In this section we apply the results in the previous sections to study
the classical nonlinear dynamics of the coupled kelvons and
quadrupole modes with $m=-2$. 
The equations of motion for the kelvons and the quadrupole mode with $m=-2$
are given by
\beq
\label{vk_eq}
i\frac{d\hat{v}_k}{dt}=E_c(k,0)q_{-2}e^{-i\delta_{-2}(k,0)t}\hat{v}_{-k}^\dagger,
\enq
\beq
i\frac{d \hat{v}_{-k}^\dagger}{dt}=-E_c(k,0)q_{-2}^* e^{i\delta_{-2}(k,0)t}\hat{v}_k,
\enq
and
\beq
\label{qm2_eq}
i\frac{d q_{-2}}{dt}=\sum_k E_c(k,0)\langle \hat{v}_k\hat{v}_{-k}\rangle
e^{i\delta_{-2}(k,0)t}.
\enq
In the above equations we assumed, an assumption we make 
throughout this paper, that the quadrupole mode is strongly excited
and can thus
be described with a complex number. In this section
we also treat the kelvons classically, and thus the operator
properties of the equations of motion will be, for the time
being, ignored.

We have shown earlier how the coupling between
the kelvons and the quadrupole mode with $m=-2$
becomes important at a certain value of the kelvon momentum $k$.
From the knowledge of the kelvon and quadrupole mode dispersions
we can solve the position of the resonance, but we still do not
know the width of this resonance. This width is 
important as it influences how many kelvon modes become occupied. 
We obtain a simple
estimate for the width of the resonance by assuming
a constant quadrupole mode amplitude $q_{-2}(0)$ and
by seeking exponential solutions for the kelvon amplitudes 
$v_k(t)\propto \exp\left(\Omega_{D,k}\,t\right)$.
The equations of motion for the kelvon amplitudes then result in an  
eigenvalue problem for the characteristic
(imaginary) frequency $\Omega_{D,k}$
\beq
\hat{\Omega}_{D,k}^2=\left(\begin{array}{cc}
(E_c(k,0)q_{-2}(0))^2 & -\delta_{-2}(k,0)E_c(k,0)q_{-2}(0) \\
-\delta_{-2}(k,0)E_c(k,0)q_{-2}(0)& (E_c(k,0)q_{-2}(0))^2 
\end{array}
\right).
\enq
This matrix has eigenvalues
\beq
\Omega_{D,k}=\pm\sqrt{E_c(k,0)q_{-2}(0)\left[E_c(k,0)q_{-2}(0)\pm |\delta_{-2}(k,0)|\right]}
\label{dyninstab}
\enq
and therefore the Kelvin modes are dynamically unstable
only when $|\delta_{-2}(k,0)|< |E_c(k,0)q_{-2}(0)|$. This 
estimate for the width of the resonance is in satisfactory agreement with
the numerical solution of the nonlinear time evolution
following from Eqs.~(\ref{vk_eq})--(\ref{qm2_eq}).
In addition, Eq.~(\ref{dyninstab}) indicates that the 
timescale for the onset of the instability is given by
$1/|E_c(k,0)q_{-2}(0)|$.

\begin{figure}
\includegraphics[width=\columnwidth]{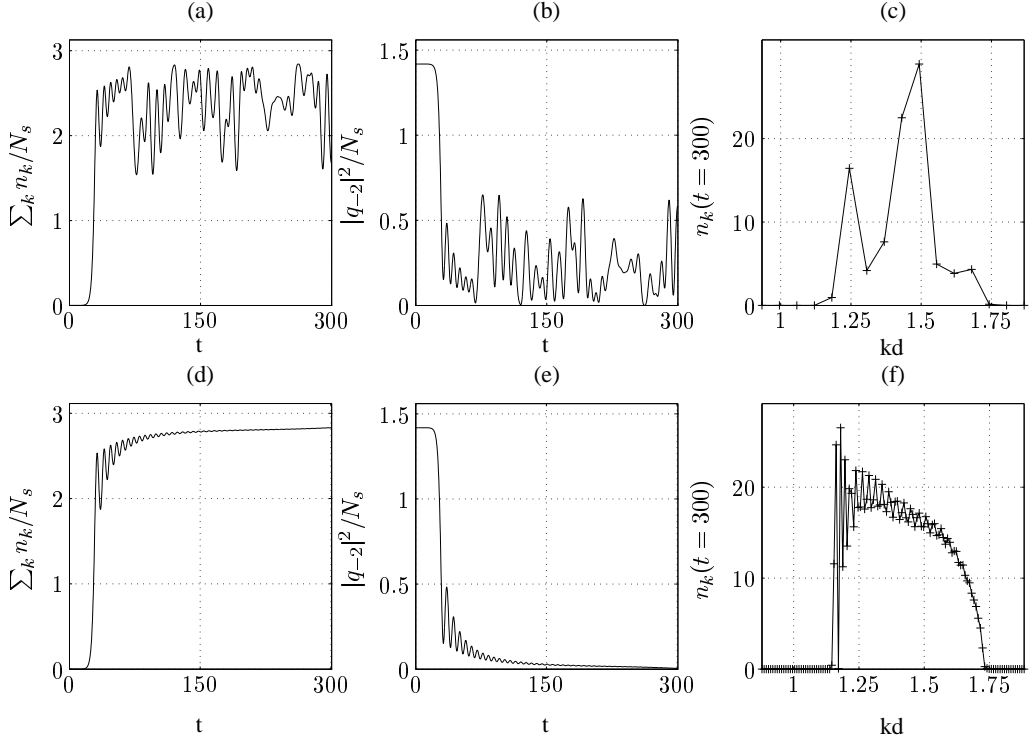}
\caption[]{(a)-(c) The time evolution in units
of $1/\omega_r$ of the total kelvon number per site,
the quadrupole density per site, and the number distribution of
kelvons in the end of the simulation, when kelvons are treated classically
and $N_s=101$.
(d)-(f) These plots used the same parameters
as the plots (a)-(c), except that the number of sites was 
larger, namely $N_s=751$.
In both sets of simulations we 
had initially $|q_{-2}|^2/N_s=\sqrt{1+2U}/10$ 
and a very small
kelvon population to act as a seed. Also, we used $J=0.1$ and $U=100$.
In plots (c) and (f) the $+$-signs indicate the actual modes
used in the numerical solution whereas the solid line is 
an interpolation between them.
\label{fig:classicalplot}}
\end{figure}

The argument above was based on the assumption of having a 
constant quadrupole field. This assumption
is bound to fail during the time evolution. When this happens,
the numerical solution of the problem becomes a necessity.
In Fig.~\ref{fig:classicalplot} we show two examples 
of the classical time evolution of the coupled system.
In this figure
only the quadrupole mode is initially strongly populated. Furthermore, 
for numerical
reasons, the kelvons have a small initial amplitude which acts as a seed
for the time evolution. When the kelvons are treated quantum mechanically
such a seed is not required as we show in the next section. 
As is clear from the figure,
the quadrupole mode is strongly damped while kelvon
modes are being populated. At longer times the system
exhibits oscillatory behaviour, but such oscillations
become weaker as the number of lattice sites is increased.

\begin{figure}
\includegraphics[width=\columnwidth]{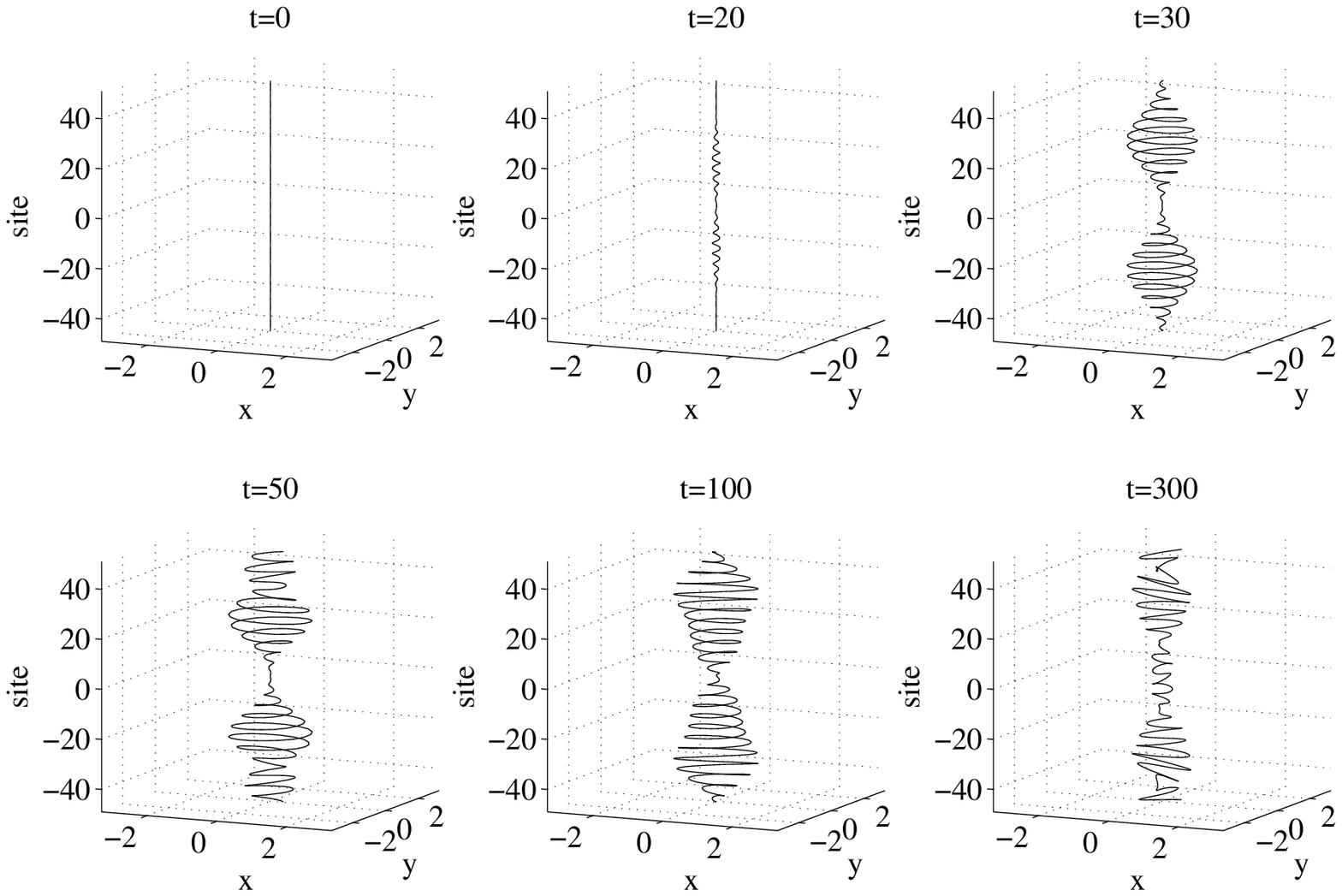}
\caption[]{Classical vortex line dynamics
when $J=0.1$, $U=100$, and $N_s=101$. The initial state
has $|q_{-2}|^2/N_s=\sqrt{1+2U}/10$ and 
a very small kelvon population to act as a seed.
The unit of time is $1/\omega_r$ and the unit of length
is $\sqrt{\hbar/m\omega_r}$.
\label{fig:classicalvortexline}}
\end{figure}

When the kelvon amplitudes are known, it is a simple matter
to calculate also the dynamics of the vortex line. In 
Fig.~\ref{fig:classicalvortexline} we show an example of 
the vortex line 
dynamics for the system solved in Fig.~\ref{fig:classicalplot}
(a)-(c). In the figure the short wavelength corresponds to
the resonant mode with the momentum $k_0$ and the
envelope with the longer wavelength is a result
of the width of the resonance.
The fact that many kelvon modes with different momenta
are occupied is also apparent from the fairly
complex structure of the vortex line.

\section{Quantum dynamics of the vortex line}
\label{sec:quantumdyn}
In this section we discuss the quantum dynamics of
the vortex line. In our numerical solution we use again
a classical quadrupole mode, like in the previous
section, but the kelvons are treated 
quantum mechanically. When the time interval $\Delta t$ is
short enough so 
that the quadrupole field $q_{-2}(t)=|q_{-2}(t)|e^{-i\phi(t)}$
can be considered as constant, we solve
the Heisenberg equations of motion for the kelvons
from time $t$ to $t+\Delta t$
analytically. The result is given by
\beq
\hat{v}_k(t+\Delta t)=\hat{v}_k(t)
\cosh\left(\Omega_c(t)\Delta t\right)-i
\hat{v}_{-k}^\dagger(t)\sinh\left(\Omega_c(t)\Delta t\right)
e^{-i\left(\phi(t)+\delta_{-2}(k,0)t\right)}
\enq
and 
\beq
\hat{v}_{-k}^\dagger(t+\Delta t)=\hat{v}_{-k}^\dagger(t)
\cosh\left(\Omega_c(t)\Delta t\right)+i
\hat{v}_{k}(t)\sinh\left(\Omega_c(t) \Delta t\right)
e^{i\left(\phi(t)+\delta_{-2}(k,0)t\right)},
\enq
where $\Omega_c(t)=E_c(k,0)|q_{-2}(t)|$. This solution enables
us to update the correlation functions needed later on. In particular,
we need the correlation functions
$\langle\hat{v}_k\hat{v}_{-k}\rangle$ and
$n_k=\langle\hat{v}_k^\dagger\hat{v}_k\rangle$
the first of which
is used to update the amplitude of the quadrupole mode.

\begin{figure}
\includegraphics[width=\columnwidth]{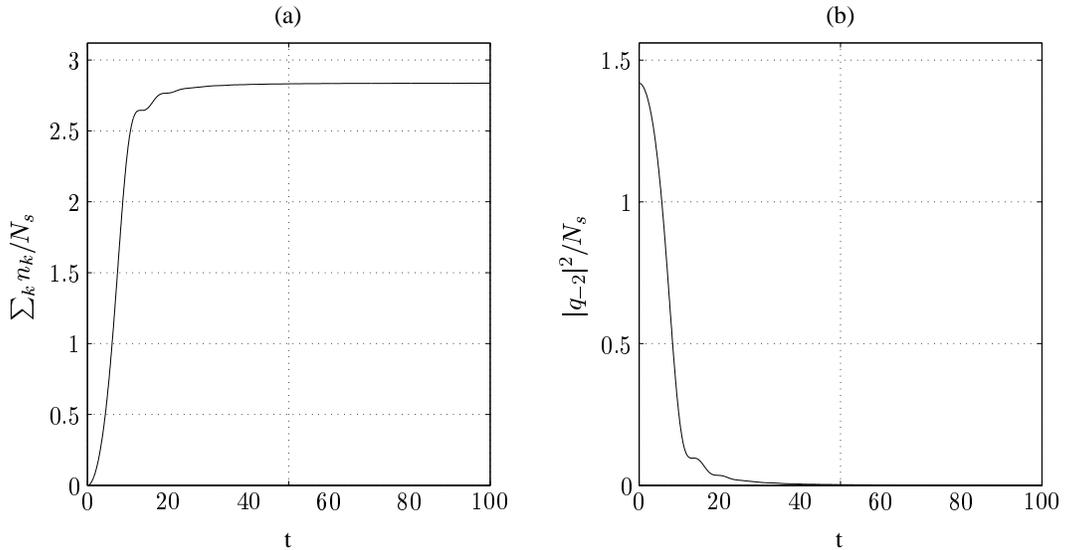}
\caption[]{(a) This plot shows the time evolution in units
of $1/\omega_r$ of the total
average kelvon number per site for quantum mechanical kelvons.
(b) This plot shows the quadrupole density per site as a function of time.
In both plots we used $J=0.1$, $U=100$, and $N_s=751$.
Initially the system was prepared
in the kelvon vacuum and with a quadrupole density
$|q_{-2}|^2/N_s=\sqrt{1+2U}/10$.
\label{fig:quantumplot}}
\end{figure}

In Figs.~\ref{fig:quantumplot} and \ref{fig:quantumplot3D} 
we show an example 
of the quantum time evolution of a system that
is initially prepared in a kelvon vacuum and with 
a nonvanishing quadrupole-mode amplitude.
It is instructive to compare 
Fig.~\ref{fig:quantumplot} with its classical counterpart in 
Fig.~\ref{fig:classicalplot} (d)-(e). The ambiguity of 
the classical initial state influences the growth rate
of the classical kelvons. Therefore, detailed comparison
with the quantum result cannot be easily made. Nevertheless, it is
clear that in the quantum mechanical simulation
the oscillatory behaviour is weaker. 
Fig.~\ref{fig:quantumplot3D} shows the kelvon-number distribution
as a function of time. After a short transient the system
settles into a nearly steady state. In principle,
revivals of the quadrupole field are possible, but the
revival time with the number of sites we used in this simulation
is longer than the length of our simulation. With
fewer sites the oscillatory behaviour becomes more 
pronounced, in agreement with the our earlier results of the 
classical simulations.

\begin{figure}
\caption[]{(Color online) This figure shows the
kelvon-number distribution as a function of time
in units of $1/\omega_r$
when the kelvons are treated fully quantum mechanically.
We used parameters
$J=0.1$, $U=100$, and $N_s=751$. Initially the system was prepared
in the kelvon vacuum and with a quadrupole density
$|q_{-2}|^2/N_s=\sqrt{1+2U}/10$.
The unit of time is $1/\omega_r$.
\label{fig:quantumplot3D}}
\end{figure}

\subsection{Transition rate into kelvons using Fermi's golden rule}
Fermi's golden rule 
\beq
\Gamma_{Q\rightarrow K}=2\pi\sum_f|\langle
\Psi_f|\hat{H}_{-2}|\Psi_i\rangle|^2\delta\left(E_f-E_i\right)
\enq
can be used to estimate the transition rate from the 
quadrupole mode into kelvons. In this expression
$|\Psi_i\rangle$ is the initial state,
$|\Psi_f\rangle$ is the final state and $E_0$ and $E_f$ are their
energies, respectively. 
In our case the initial state is a kelvon vacuum with a coherent
quadrupole mode of amplitude $q_{-2}(0)$, i.e.,
$|\Psi_i\rangle=|q_{-2}(0),\;n_{k_0}=0,\;n_{-k_0}=0\rangle$.
By calculating the matrix element 
$\langle\Psi_f|\hat{H}_{-2}|\Psi_i\rangle$
and the density of states  we obtain
\beq
\label{eq_fermi}
\Gamma_{Q\rightarrow K}=
\frac{\pi E_c(k_0,0)^2|q_{-2}\left(0\right)|^2}{2J\Gamma[0,B_0^2]|\sin\left(k_0d\right)|},
\enq
where $k_0$ is the resonant value of the kelvon momentum and 
$d=\lambda/2$ is the lattice spacing.

\begin{figure}
\includegraphics[width=\columnwidth]{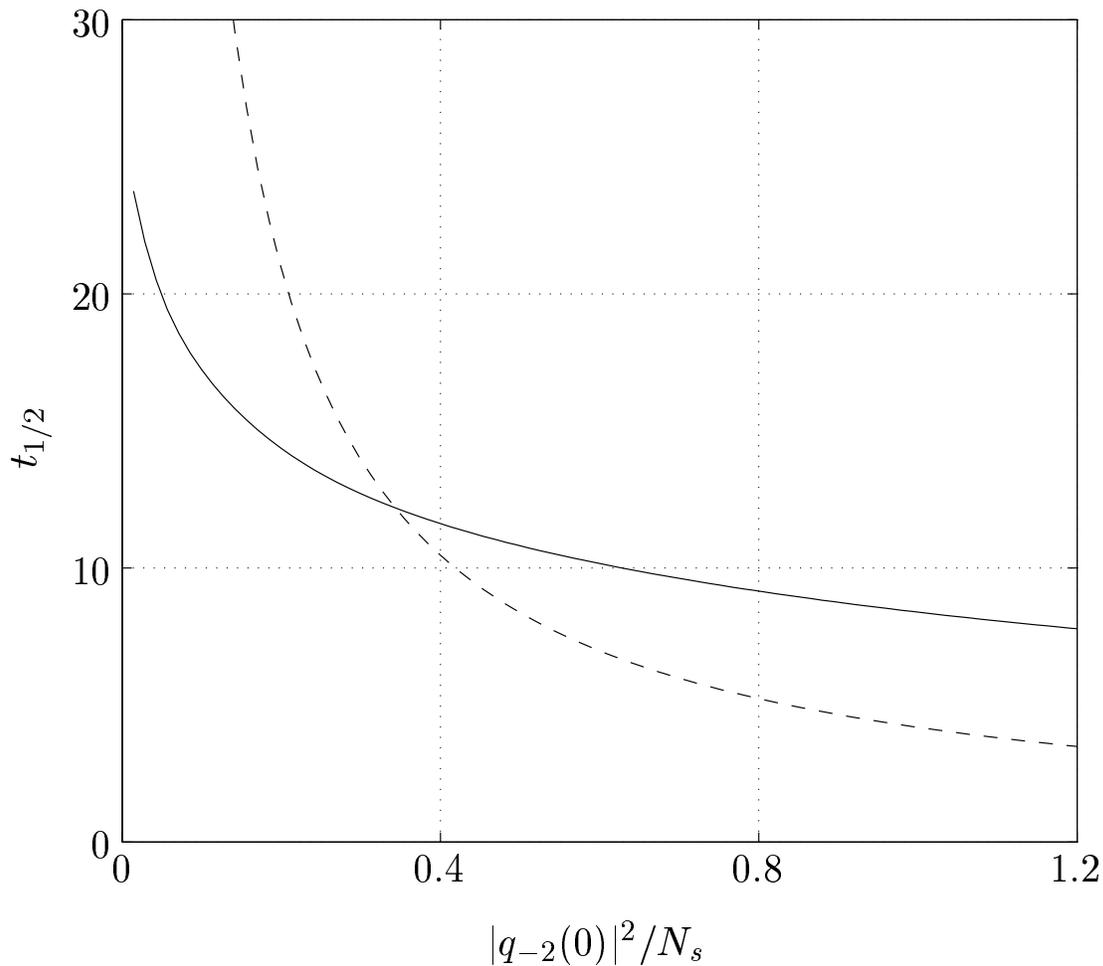}
\caption[]{This figure shows the time it takes for the
quadrupole field density to halve from its initial value
$|q_{-2}(0)|^2$. 
The solid line is based on the numerical solution
of the quantum multimode problem, while the dashed line
is based on Fermi's golden rule.
We used parameters
$J=0.1$, $U=100$, $N_s=751$, and the unit of time is
$1/\omega_r$.
\label{fig:Halflife}}
\end{figure}

Fig.~\ref{fig:Halflife} compares the Fermi's golden rule prediction
with the
numerical solution of the quantum multimode problem by presenting
the half-life of the quadrupole field as a function of initial
quadrupole field strength. As can be seen
from this figure, Fermi's golden rule underestimates
the decay rate for narrow resonances (small values of the 
initial quadrupole field) while overestimating the decay
rate for wide resonances. The exact solution deviates from the
Fermi's golden rule result, since  Fermi's golden rule
ignores the finite lifetime of the final states. In particular, 
the detunings in Eq.~(\ref{qm2_eq}) contribute
to the phase factors which determine whether the quadrupole
mode is increasing or decreasing. At small times the quadrupole mode
is always depleted, but at longer times the behaviour depends
on the width of the resonance. 

\section{Nonequilibrium squeezing}
\label{sec:squeezing}
As the interaction between the quadrupole mode with $m=-2$
and the kelvons closely resembles the squeezing Hamiltonians
in quantum optics, it is worth while to investigate
the squeezing of the vortex
in detail. This will be done in the following two
subsections. We start
in Sec.~\ref{sec:singlemodesqueezing} by
discussing the squeezing when the quadrupole mode
is resonantly coupled to only one Kelvin mode. Understanding this
important special case enables us
to explore, in Sec.~\ref{sec:multimodesqueezing},
the experimentally more realistic problem of a multimode
squeezing.

\subsection{Single-mode squeezing}
\label{sec:singlemodesqueezing}
In the simplest case we have only one resonant kelvon mode
at momentum $k_0$. For simplicity we assume that initially the quadrupole 
mode is a real number. We can then define two quadrature operators
\beq
\hat{b}_1
=\frac{1}{2\sqrt{2}}\left[i\hat{v}_{-k_0}-i\hat{v}_{-k_0}^\dagger
-\hat{v}_{k_0}-\hat{v}_{k_0}^\dagger\right]
\label{eq_b1}
\enq
and
\beq
\hat{b}_2=\frac{1}{2\sqrt{2}}\left[\hat{v}_{-k_0}+i\hat{v}_{k_0}
+\hat{v}_{-k_0}^\dagger-i\hat{v}_{k_0}^\dagger\right].
\label{eq_b2}
\enq
These operators obey an uncertainty relation
\beq
\sqrt{\langle \hat{b}_1^2\rangle}\sqrt{\langle \hat{b}_1^2\rangle}\le
\frac{1}{4}
\enq
and their commutator is 
\beq
\left[\hat{b}_1,\hat{b}_2\right]=\frac{i}{2}.
\enq
These operators are analogous to the 
squeezing operators familiar from quantum optics~\cite{Scully1997a}, 
but are somewhat
counter intuitive. We gain more insight when we realize that
the operators $\hat{b}_i$ are related to the position
operators $\hat{x}_{k_0}(\theta)$ and $\hat{y}_{k_0}(\theta)$ 
in a coordinate system rotated by an angle $\theta$.
In particular, we have
$\hat{b}_1=\sqrt{NB_0}\,\hat{x}_{k_0}(\theta)$ and 
$\hat{b}_2=\sqrt{NB_0}\,\hat{y}_{k_0}(\theta)$. 
The rotation angle $\theta$ of the vortex position operators
depends on the initial phase of the quadrupole mode,
because Eqs.~(\ref{eq_b1}) and (\ref{eq_b2})
are in principle modified if $q_{-2}(0)$ is not real.
If the quadrupole mode is initially $|q_{-2}(0)|e^{i\nu_Q}$
the rotation angle is given by 
\beq
\theta=\left(-\frac{3\pi}{2}+\nu_Q\right)/2.
\enq
The proportionality of the quadrature operators to the
vortex position operators
implies that the possible squeezing of the vortex state
is reflected in the corresponding deformation of the
vortex-position distribution.

\begin{figure}
\includegraphics[width=\columnwidth]{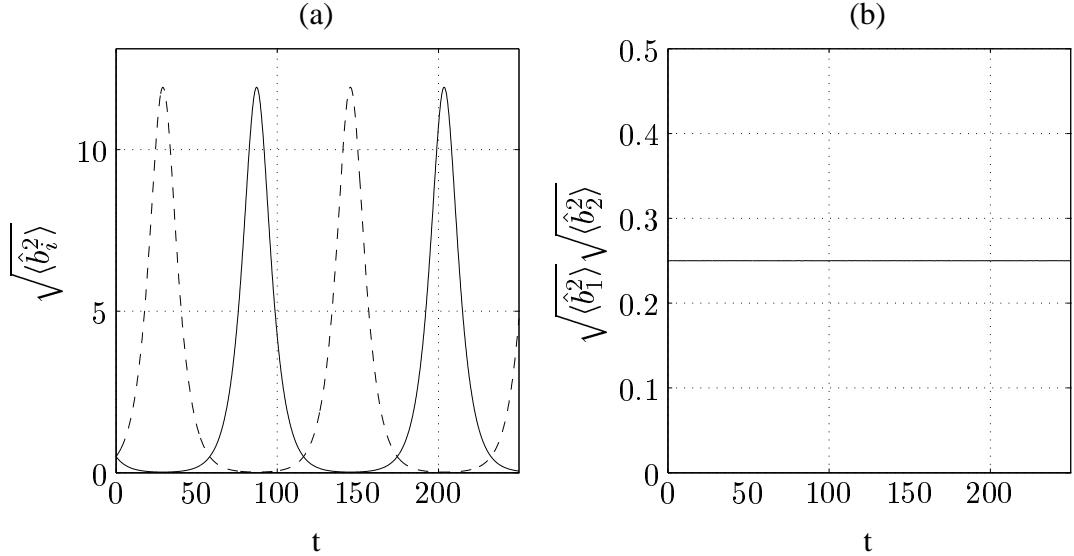}
\caption[]{(a) The time evolution in units of 
$1/\omega_r$ of the quadrature operator uncertainty
$\sqrt{\langle \hat{b}_i^2\rangle}$. 
We used parameters
$N_s=100$, $N=100$, $J=0.1$, $U=100$, and 
$|q_{-2}(0)|^2/N_s=\sqrt{1+2U}/10$.
The solid line is for the operator $\hat{b}_1$ 
and the dashed line for the operator $\hat{b}_2$. 
(b) The figure shows the time evolution
for the product 
$\sqrt{\langle \hat{b}_1^2\rangle}\sqrt{\langle \hat{b}_2^2\rangle}$.
\label{fig:singlemode_squeeze}}
\end{figure}

In Fig.~\ref{fig:singlemode_squeeze} we show an example of the time evolution
of the squares of the quadrature operators as well as the product
$
\sqrt{\langle \hat{b}_1^2\rangle}\sqrt{\langle \hat{b}_2^2\rangle}.
$
As can be seen from this figure, the minimum uncertainty state
(the vacuum) remains a minimum uncertainty state, but can become 
strongly squeezed during the time evolution.
In coordinate space this squeezing is reflected in
the deformation of an initially circular uncertainty ellipse
into an ellipse with dramatically different 
main axes. Furthermore, since the initial quadrupole mode amplitude
was choosen as a real number, the main axes are always
along the lines $y=\pm x$.

\subsection{Multimode squeezing}
\label{sec:multimodesqueezing}
The main difference between the multimode squeezing, as
opposed to the resonant single-mode squeezing, is that the
area of the uncertainty ellipse is no longer constant.
This means that when
the time evolution starts from the kelvon vacuum, the
state evolves away from the minimum uncertainty state. 

Generalizing our earlier results for the
single-mode squeezing we define the multimode 
quadrature operators in terms
of single-mode quadrature operators for kelvons with momentum $k$, 
 $\hat{b}_{i,k}$, as
\beq
b_i=\frac{1}{N_s}\sum_k \hat{b}_{i,k}.
\enq
For these operators the commutator is again
$\left[\hat{b}_1,\hat{b}_2\right]=\frac{i}{2}$. However, rather than 
embarking on this more formal road, we choose to focus on the shape of the
vortex-position distribution. This distribution is
measurable and provides the necessary signatures of the
squeezing.

When the initial state of the vortex line is a kelvon vacuum, the fluctuations
of the vortex position 
in a coordinate system rotated by an angle $\theta$
are given by
\begin{eqnarray}
\langle \hat{x}(\theta,t)^2\rangle&=&
\frac{1}{2NN_sB_0}\left[1+2n_0(t)+2|\langle v_0(t)v_0(t)\rangle|
\sin\left(2\theta+\phi_0(t)\right)
\right.\nonumber\\
&+&\left.\sum_{k\neq 0}\left(n_k(t)+\frac{1}{2}+|\langle v_k(t)v_{-k}(t)\rangle|
\sin\left(2\theta+\phi_k(t)\right)
\right)\right]
\end{eqnarray}
and
\begin{eqnarray}
\langle \hat{y}(\theta,t)^2\rangle&=&
\frac{1}{2NN_sB_0}\left[1+2n_0(t)-2|\langle v_0(t)v_0(t)\rangle|
\sin\left(2\theta+\phi_0(t)\right)
\right.\nonumber\\
&+&\left.\sum_{k\neq 0}\left(n_k(t)+\frac{1}{2}-|\langle v_k(t)v_{-k}(t)\rangle|
\sin\left(2\theta+\phi_k(t)\right)
\right)
\right],
\end{eqnarray}
where we defined $\langle v_{k}(t)v_{-k}(t)\rangle=
|\langle v_{k}(t)v_{-k}(t)\rangle|e^{i\phi_k(t)}$. 
For this initial state the quantities $\langle \hat{x}(\theta,t)^2\rangle$
and $\langle \hat{y}(\theta,t)^2\rangle$
are independent of the site. Furthermore, 
their maximum and minimum with respect
to the rotation angle define the lengths  
$\sigma_x^2(t)={\rm max}\left[\langle \hat{x}(\theta,t)^2\rangle\right]$ and 
$\sigma_y^2(t)={\rm min}\left[\langle \hat{y}(\theta,t)^2\rangle\right]$ 
of the main axes of the uncertainty ellipse
as well as their directions $\theta_{max}(t)$ and $\theta_{max}(t)+\pi/2$. 
In terms of these
quantities and in a coordinate system rotated by an angle $\theta_{max}(t)$, 
the uncertainty ellipse is defined by
\beq
\frac{x^2}{\sigma_x^2(t)}+\frac{y^2}{\sigma_y^2(t)}=1.
\enq

The main axes of the uncertainty ellipse are in general time dependent. However,
we can define the deformation 
parameter $\epsilon(t)$ of the uncertainty ellipse 
by changing the rotation angle $\theta(t)$ of the coordinate system
as a funtion of time.
In this way we define the deformation parameter as 
\beq
\epsilon(t)=\frac{\sigma_x^2(t)-\sigma_y^2(t)}
{\sigma_x^2(t)+\sigma_y^2(t)}.
\enq

\begin{figure}
\includegraphics[width=\columnwidth]{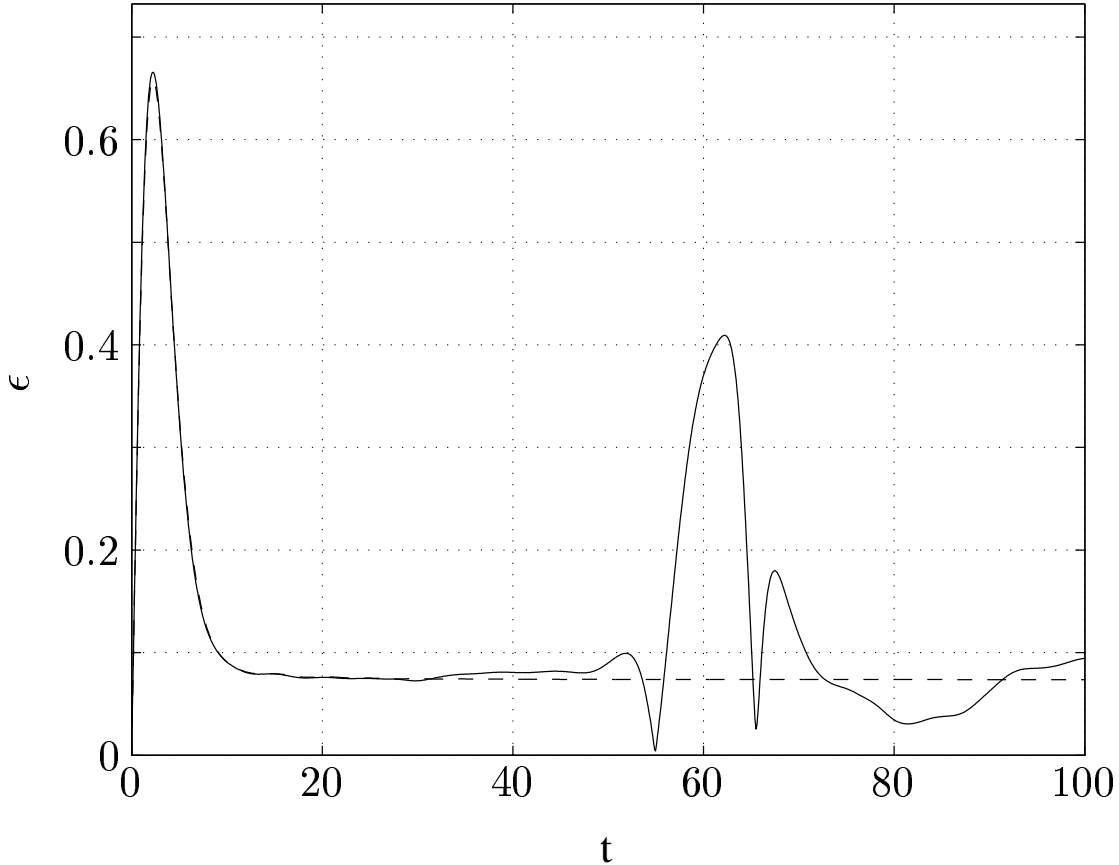}
\caption[]{This figure shows the deformation parameter
of the uncertainty ellipse with two different 
number of lattice sites. The solid line was calculated with
$N_s=101$ and the dashed line with $N_s=751$. In addition,
we used parameters $J=0.1$, $U=100$, and 
$|q_{-2}(0)|^2/N_s=\sqrt{1+2U}/10$. The unit of time is
$1/\omega_r$.
\label{fig:epsilonplot}}
\end{figure}

\begin{figure}
\includegraphics[width=\columnwidth]{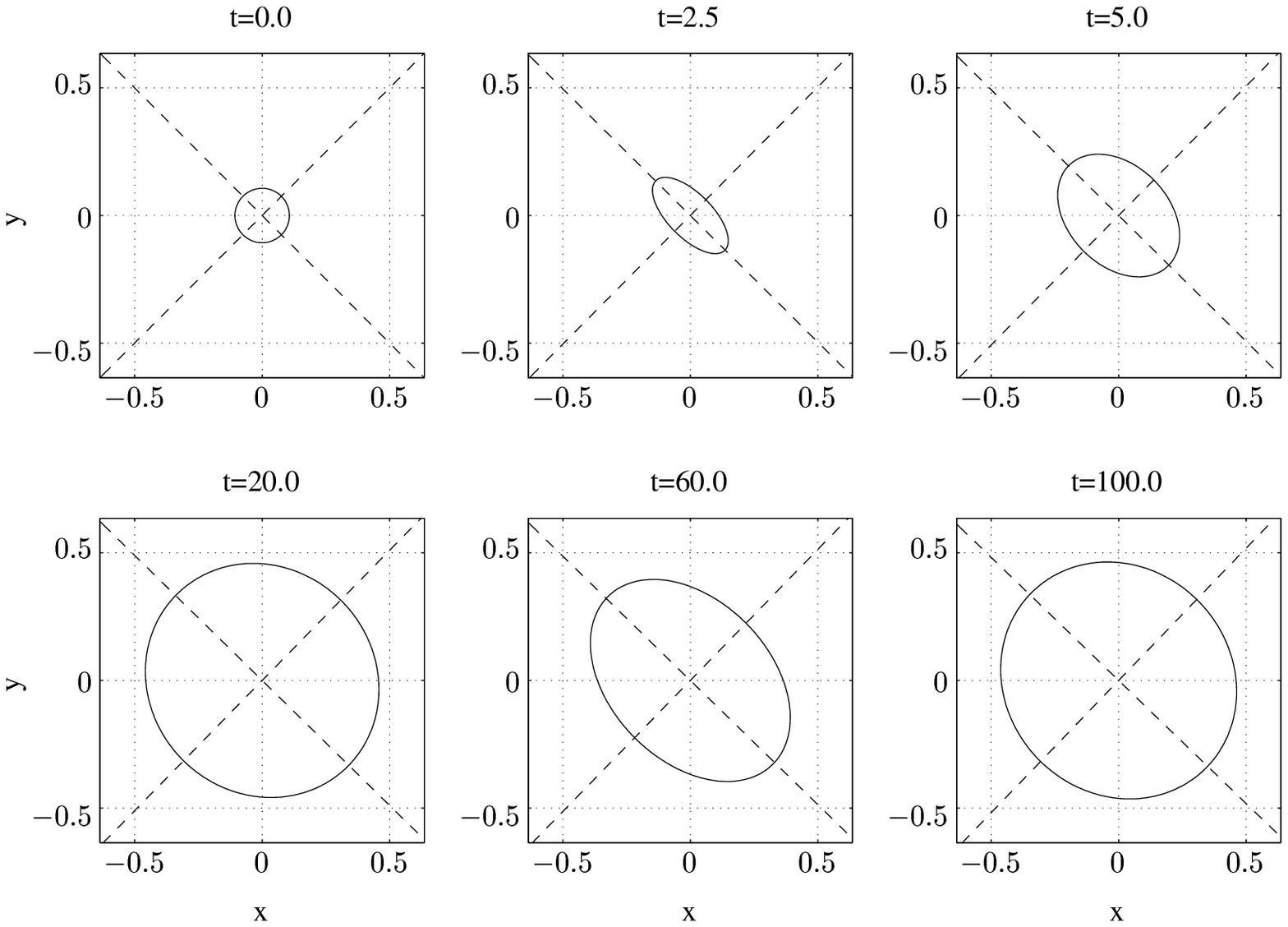}
\caption[]{This figure shows snapshots of the 
time evolution in units of $1/\omega_r$
of the uncertainty ellipse when $N_s=101$. 
In addition, we used parameters $J=0.1$, $U=100$, and 
$|q_{-2}(0)|^2/N_s=\sqrt{1+2U}/10$. Dashed lines indicate the directions
of the main axes of the uncertainty ellipse.
The unit of time is $1/\omega_r$ and the unit of length
is $\sqrt{\hbar/m\omega_r}$.
\label{fig:uncertNs101}}
\end{figure}

\begin{figure}
\includegraphics[width=\columnwidth]{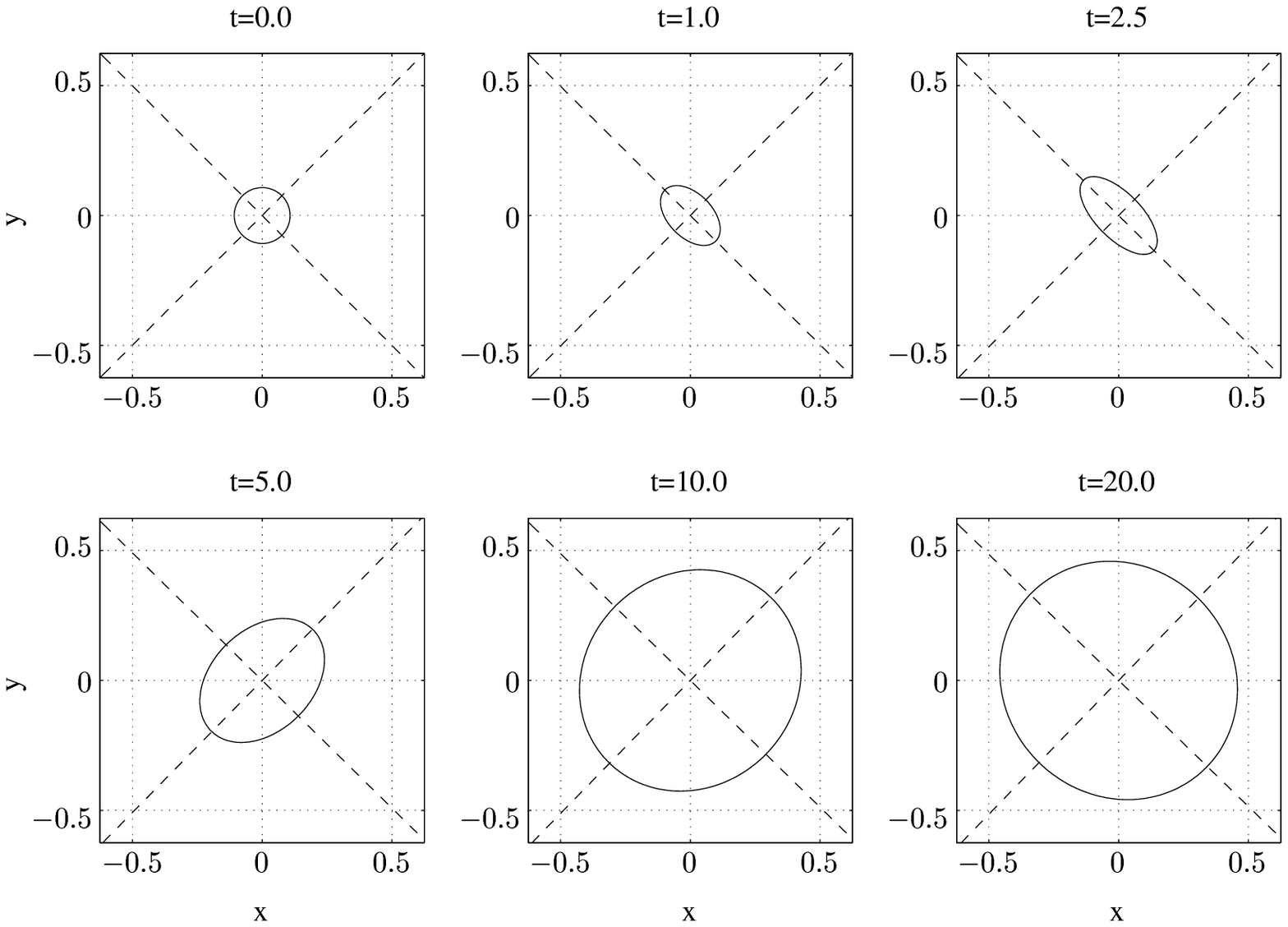}
\caption[]{This figure shows snapshots of the 
time evolution of the uncertainty ellipse with $N_s=751$. 
In addition, we used parameters $J=0.1$, $U=100$, and 
$|q_{-2}(0)|^2/N_s=\sqrt{1+2U}/10$. Dashed lines indicate the directions
of the main axes of the uncertainty ellipse.
The unit of time is $1/\omega_r$ and the unit of length
is $\sqrt{\hbar/m\omega_r}$.
\label{fig:uncertNs751}}
\end{figure}

In Fig.~\ref{fig:epsilonplot}
we plot the time evolution of the deformation parameter with
two different number of lattice sites. This figure demonstrates
how the uncertainty ellipse first becomes strongly
deformed and then settles into a more symmetric configuration.
In the simulation with fewer number of lattice sites
the strong deformation of the uncertainty ellipse revives
at about $t=60$. With larger number of lattice sites the
behaviour is, at this timescale, more irreversible.
In Figs.~\ref{fig:uncertNs101} and~\ref{fig:uncertNs751}
we show snapshots of the uncertainty ellipses corresponding to the
simulations in Fig.~\ref{fig:epsilonplot}. In these figures
we also indicate the main axes of the ellipses. The direction of the
main axes depend on the initial phase of the quadrupole mode.

\subsection{Experimental observation}
\label{sec:experimentalsqueeze}
As noted in the previous subsection the fluctuations
$\langle \hat{x}\left(\theta\right)^2\rangle$ and $\langle \hat{y}\left(\theta\right)^2\rangle$
are independent of the site. Therefore, in each site the fluctuations of the vortex
position are described by the same uncertainty ellipse and the measurement
of the vortex position in each site sample the same distribution.
However, the number of samples must be large enough to 
distinguish the squeezing from statistical fluctuations. 
The {\it in situ} imagining of the three-dimensional vortex line is difficult, but not
impossible~\cite{Bretin2003a,Anderson2001a}. 


In many experiments the trap potential is turned off and the system
is allowed to expand. This makes the experimental imaging of the condensate as well
as the vortex easier. Therefore, it is important to study
how the vortex-position distribution evolves under 
the expansion~\cite{Castin1996a,Kagan1996a,Lundh1998a}.
We assume that only the magnetic trap is turned off
and that the optical potential is left unchanged. When the small
inter-site coupling is ignored, each site will expand
independently in two-dimensions. 

We study the expansion using
the same variational ansatz we have used until now except 
that the condensate size parameter $B_0$ is replaced with a
time-dependent complex variational parameter 
$B_0\rightarrow B_0'(t)+iB_0''(t)$. The vortex core size changes as the
condensate expands and we take this into account by using 
a time-dependent short distance cut-off
\beq
\xi(t)=\frac{1}{\sqrt{2UB_0'(t)}}.
\label{eq:xit}
\enq
This expression corresponds to the aproximation that the 
vortex-core size is always equal to the coherence
length in the center of the condensate. At $t=0$, before
the condensate expands, Eq.~(\ref{eq:xit})
agrees with our earlier equilibrium
value for the short distance cut-off in the strongly-interacting
Thomas-Fermi limit because $B_0'(0)\approx 1/\sqrt{2U}$.

Up to second order in $B_0'(t)$
The equations of motion for the variational parameters
are
\beq
\dot{x}_n(t)=C(t)x_n(t)+\omega_0(t) y_n(t),
\enq
\beq
\dot{y}_n(t)=C(t)y_n(t)-\omega_0(t) x_n(t),
\enq
\beq
\dot{B}_0'(t)=2B_0''(t)\left[1-\frac{1}{8U^2}\right]B_0'(t),
\enq
and
\beq
\dot{B}_0''(t)=B_0''(t)^2\left[1-\frac{1}{8U^2}\right]
+B_0'(t)^2\left[\gamma-1-\ln 2-2U-\ln U+\frac{1}{2U}-\frac{5}{16U^2}\right].
\enq
To obtain the last equation we also expanded up
to second order in $1/U^2$. Furthermore, $\gamma\approx 0.577216$ 
is the Euler-Mascheroni constant,
\beq
C(t)=-B_0''(t)\left[1-\frac{1}{8U^2}\right],
\enq
and 
\beq
\omega_0(t)=\Omega+\frac{B_0'(t)}{2}\left(1-\Gamma\left[0,B_0'(t)^2\right]\right).
\enq
Importantly, 
from these equations we see that the dynamics of the expansion is independent
of the vortex position. The real functions $C(t)$ and $\omega_0(t)$
depend on $B_0''(t)$ and $B_0'(t)$, respectively, but since 
the expansion is independent of the vortex 
dynamics they can be considered as, in principle, given functions. 
The equations of motion for the vortex positions are then solved by
\beq
x_n(t)=\exp\left(\int_0^t C(t)dt\right)\left[
x_n(0)\cos\left(\int_0^t \omega_0(t)dt\right)
+y_n(0)\sin\left(\int_0^t \omega_0(t)dt\right)
\right]
\enq
and
\beq
y_n(t)=\exp\left(\int_0^t C(t)dt\right)\left[
y_n(0)\cos\left(\int_0^t \omega_0(t)dt\right)
-x_n(0)\sin\left(\int_0^t \omega_0(t)dt\right)
\right].
\enq
This result shows that the vortex dynamics during
the condensate expansion is a combination of
precession and scaling. The precession angle and the scaling 
factor only depend on the dynamics of the condensate expansion
and are the same irrespective of where the vortex was initially
located. Therefore, a squeezed vortex distribution 
expands, to a very good accuracy, homologously,
i.e., without changing its aspect ratio. 
The squeezed distribution is therefore observable even after 
expansion of the condensate.

\begin{figure}
\includegraphics[width=\columnwidth]{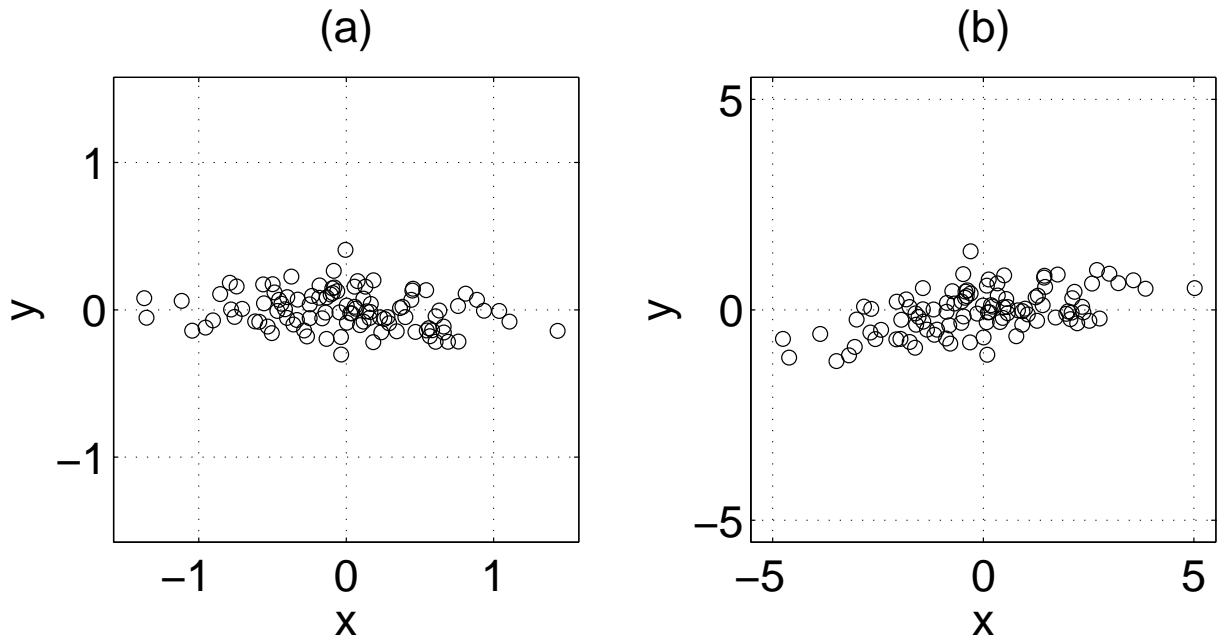}
\caption[]{This figure shows the vortex-position distribution
in units of $\sqrt{\hbar/m\omega_r}$
(a) before and (b) after the expansion where condensate size increased 
by a factor of $3.5$. We used parameters $J=0.1$, $U=100$, 
$\Omega=0$, and $N_s=101$. Initially vortex positions were sampled from
a two-dimensional Gaussian distribution with an asymmetry
of $\epsilon=0.9$ and the product of the widths was
$\sigma_x\sigma_y=B_0$.
\label{fig:expansionplot}}
\end{figure}

In Fig.~\ref{fig:expansionplot} we demonstrate how
the expansion changes the squeezed vortex-position distribution 
by solving the equations of motion
explicitly. This figure makes the homologuous character
of the expansion clearly apparent.

\section{Summary and conclusions}
\label{sec:conclusions}
We derived the quantum theory of the vortex line in a
one-dimensional optical lattice and obtained the dispersion
relation of the Kelvin modes. Our variational approach 
enabled us to also include quadrupole modes into the system
and study their coupling with the kelvons.
The coupling turned out to be described by the squeezing Hamiltonian and 
we solved the dynamics of the vortex line coupled with the
quadrupole mode both classically as well as quantum mechanically.
Furthermore, we found that the vortex line can indeed be squeezed 
in this system if we drive the quadrupole modes
sufficiently strongly. We also discussed 
some of the expected experimental signatures.

Our theory can be applied to a variety of new problems.
For example, exploring the properties of the vortex line
at finite temperatures is possible by adapting the theory we 
have used in this paper~\cite{Martikainen2003d}. Also, 
the interplay between the superfluid Mott-insulator
transition, the quantum melting of 
a vortex lattice, and the quantum-Hall regime in an optical lattice
is a very interesting topic for further research.

\begin{acknowledgments}
We thank Randy Hulet for suggesting to us
to consider also the vortex-position distribution
after expansion.
This work is supported by the Stichting voor Fundamenteel Onderzoek der 
Materie (FOM) and by the Nederlandse Organisatie voor 
Wetenschaplijk Onderzoek (NWO).
\end{acknowledgments}

\bibliographystyle{apsrev}

\end{document}